\documentclass[aps,amsmath,preprint,epsf,superscriptaddress,nofootinbib,notitlepage]{revtex4-1}
\usepackage{graphicx}
\usepackage{bm}
\usepackage{color}
\usepackage{amsmath,amssymb}	
\usepackage{hyperref}
\usepackage{setspace}
\usepackage{longtable}
\usepackage{booktabs}
\usepackage{cancel}

\pdfoutput=1

\def\slashchar#1{\setbox0=\hbox{$#1$} % set a box for #1
\dimen0=\wd0 % and get its size
\setbox1=\hbox{/} \dimen1=\wd1 % get size of /
\ifdim\dimen0>\dimen1 % #1 is bigger
\rlap{\hbox to \dimen0{\hfil/\hfil}} % so center / in box
#1 % and print #1
\else % / is bigger
\rlap{\hbox to \dimen1{\hfil$#1$\hfil}} % so center #1
/ % and print /
\fi}

\begin{document}
%%%%%%%%%%%%%%%%%  Defs. %%%%%%%%%%%%%%%%%%%%%%%%%%%%%%
\def\a{\alpha}
\def\b{\beta}
\def\c{\varepsilon}
\def\d{\delta}
\def\e{\epsilon}
\def\f{\phi}
\def\g{\gamma}
\def\h{\theta}
\def\k{\kappa}
\def\l{\lambda}
\def\m{\mu}
\def\n{\nu}
\def\p{\psi}
\def\q{\partial}
\def\r{\rho}
\def\s{\sigma}
\def\t{\tau}
\def\u{\upsilon}
\def\v{\varphi}
\def\w{\omega}
\def\x{\xi}
\def\y{\eta}
\def\z{\zeta}
\def\D{{\mit \Delta}}
\def\G{\Gamma}
\def\H{\Theta}
\def\L{\Lambda}
\def\F{\Phi}
\def\P{\Psi}

\def\S{\Sigma}

\def\o{\over}
\def\beq{\begin{eqnarray}}
\def\eeq{\end{eqnarray}}
\newcommand{\gsim}{ \mathop{}_{\textstyle \sim}^{\textstyle >} }
\newcommand{\lsim}{ \mathop{}_{\textstyle \sim}^{\textstyle <} }
\newcommand{\vev}[1]{ \left\langle {#1} \right\rangle }
\newcommand{\bra}[1]{ \langle {#1} | }
\newcommand{\ket}[1]{ | {#1} \rangle }
\newcommand{\EV}{ {\rm eV} }
\newcommand{\KEV}{ {\rm keV} }
\newcommand{\MEV}{ {\rm MeV} }
\newcommand{\GEV}{ {\rm GeV} }
\newcommand{\TEV}{ {\rm TeV} }
\def\diag{\mathop{\rm diag}\nolimits}
\def\Spin{\mathop{\rm Spin}}
\def\SO{\mathop{\rm SO}}
\def\O{\mathop{\rm O}}
\def\SU{\mathop{\rm SU}}
\def\U{\mathop{\rm U}}
\def\Sp{\mathop{\rm Sp}}
\def\SL{\mathop{\rm SL}}
\def\tr{\mathop{\rm tr}}

\def\IJMP{Int.~J.~Mod.~Phys. }
\def\MPL{Mod.~Phys.~Lett. }
\def\NP{Nucl.~Phys. }
\def\PL{Phys.~Lett. }
\def\PR{Phys.~Rev. }
\def\PRL{Phys.~Rev.~Lett. }
\def\PTP{Prog.~Theor.~Phys. }
\def\ZP{Z.~Phys. }

% draw box with width #1 pt and line thickness #2 pt
\newcommand{\drawsquare}[2]{\hbox{%
\rule{#2pt}{#1pt}\hskip-#2pt%  left vertical
\rule{#1pt}{#2pt}\hskip-#1pt%  lower horizontal
\rule[#1pt]{#1pt}{#2pt}}\rule[#1pt]{#2pt}{#2pt}\hskip-#2pt%  upper horizontal
\rule{#2pt}{#1pt}}% right vertical

\def\vbr{\vphantom{\sqrt{F_e^i}}}% vertical brace for tables
% Young tableaux
\newcommand{\fund}{\drawsquare{6.5}{0.4}}%  fundamental
\newcommand{\afund}{\overline{\fund}}
\newcommand{\symm}{\drawsquare{6.5}{0.4}\hskip-0.4pt%
        \drawsquare{6.5}{0.4}}%  symmetric second rank tensor
\newcommand{\asymm}{\raisebox{-3pt}{\drawsquare{6.5}{0.4}\hskip-6.9pt%
        \raisebox{6.5pt}{\drawsquare{6.5}{0.4}}}}%  antisymmetric second rank
\newcommand{\asymmthree}{\raisebox{-7pt}{\drawsquare{6.5}{0.4}}\hskip-6.9pt%
\raisebox{-0.5pt}{\drawsquare{6.5}{0.4}}\hskip-6.9pt%
\raisebox{6pt}{\drawsquare{6.5}{0.4}}}% antisymmetric third rank
\newcommand{\asymmfour}{\raisebox{-10pt}{\drawsquare{6.5}{0.4}}\hskip-6.9pt%
\raisebox{-3.5pt}{\drawsquare{6.5}{0.4}}\hskip-6.9pt%
\raisebox{3pt}{\drawsquare{6.5}{0.4}}\hskip-6.9pt%
        \raisebox{9.5pt}{\drawsquare{6.5}{0.4}}}%  antisymmetric fourth rank
\newcommand{\Ythrees}{\raisebox{-.5pt}{\drawsquare{6.5}{0.4}}\hskip-0.4pt%
          \raisebox{-.5pt}{\drawsquare{6.5}{0.4}}\hskip-0.4pt% 
          \raisebox{-.5pt}{\drawsquare{6.5}{0.4}}}%  symmetric third rank
\newcommand{\Yfours}{\raisebox{-.5pt}{\drawsquare{6.5}{0.4}}\hskip-0.4pt%
          \raisebox{-.5pt}{\drawsquare{6.5}{0.4}}\hskip-0.4pt% 
          \raisebox{-.5pt}{\drawsquare{6.5}{0.4}}\hskip-0.4pt% 
          \raisebox{-.5pt}{\drawsquare{6.5}{0.4}}}%  symmetric fourth rank
\newcommand{\Ythreea}{\raisebox{-3.5pt}{\drawsquare{6.5}{0.4}}\hskip-6.9pt%
        \raisebox{3pt}{\drawsquare{6.5}{0.4}}\hskip-6.9pt
        \raisebox{9.5pt}{\drawsquare{6.5}{0.4}}}
\newcommand{\Yfoura}{\raisebox{-3.5pt}{\drawsquare{6.5}{0.4}}\hskip-6.9pt%
        \raisebox{3pt}{\drawsquare{6.5}{0.4}}\hskip-6.9pt
        \raisebox{9.5pt}{\drawsquare{6.5}{0.4}}\hskip-6.9pt
        \raisebox{16pt}{\drawsquare{6.5}{0.4}}}
\newcommand{\Yadjoint}{\raisebox{-3.5pt}{\drawsquare{6.5}{0.4}}\hskip-6.9pt%
        \raisebox{3pt}{\drawsquare{6.5}{0.4}}\hskip-0.4pt
        \raisebox{3pt}{\drawsquare{6.5}{0.4}}}%  SU(3) adjoint
\newcommand{\Ysquare}{\raisebox{-3.5pt}{\drawsquare{6.5}{0.4}}\hskip-0.4pt%
        \raisebox{-3.5pt}{\drawsquare{6.5}{0.4}}\hskip-13.4pt%
        \raisebox{3pt}{\drawsquare{6.5}{0.4}}\hskip-0.4pt%
        \raisebox{3pt}{\drawsquare{6.5}{0.4}}}%  4 boxes in a square
\newcommand{\Yflavor}{\Yfund + \overline{\Yfund}} % box anti-box pair
\newcommand{\Yoneoone}{\raisebox{-3.5pt}{\drawsquare{6.5}{0.4}}\hskip-6.9pt%
        \raisebox{3pt}{\drawsquare{6.5}{0.4}}\hskip-6.9pt%
        \raisebox{9.5pt}{\drawsquare{6.5}{0.4}}\hskip-0.4pt%
        \raisebox{9.5pt}{\drawsquare{6.5}{0.4}}}%

%%%%%%%%%%%%%%%%%%%%%%%%%%%%%%%%%%
%%%%%%%%%%% Title page %%%%%%%%%%%
%%%%%%%%%%%%%%%%%%%%%%%%%%%%%%%%%%

%\bigskip

\preprint{IPMU17-0021}

\title{Dark Matter Candidates in a Visible Heavy QCD Axion Model}
\author{Hajime Fukuda}
\email[e-mail: ]{hajime.fukuda@ipmu.jp}
\affiliation{Kavli IPMU (WPI), UTIAS, The University of Tokyo, Kashiwa, Chiba 277-8583, Japan}

\author{Masahiro Ibe}
\email[e-mail: ]{ibe@icrr.u-tokyo.ac.jp}
\affiliation{Kavli IPMU (WPI), UTIAS, The University of Tokyo, Kashiwa, Chiba 277-8583, Japan}
\affiliation{ICRR, The University of Tokyo, Kashiwa, Chiba 277-8582, Japan}

\author{Tsutomu T. Yanagida}
\email[e-mail: ]{tsutomu.tyanagida@ipmu.jp}
\affiliation{Kavli IPMU (WPI), UTIAS, The University of Tokyo, Kashiwa, Chiba 277-8583, Japan}

\date{\today}
\begin{abstract}
In this paper, we discuss dark matter candidates in a visible heavy QCD axion model. 
There, a mirror copied sector of the Standard Model
with mass scales larger than the Standard Model is introduced.
By larger mass scales of the mirrored sector, the QCD axion is made heavy 
via the axial anomaly in the mirrored sector without spoiling the Peccei-Quinn mechanism to solve the strong $CP$-problem.
Since the mirror copied sector possesses the same symmetry structure with the Standard Model sector,
the model predicts multiple stable particles.
As we will show, the mirrored charged pion and the mirrored electron 
can be  viable candidates for dark matter.
They serve as self-interacting dark matter with a long range force.
We also show that the mirrored neutron can be lighter than the mirrored proton in a certain parameter region.
There, the mirrored neutron can also be a viable dark matter candidate when its mass is around $100$\,TeV.
It is also shown that the mirrored neutrino can also be a viable candidate for dark matter.
\end{abstract}
\maketitle

\newpage

%%%%%%%%%%%%%%%%%%%%%%%%%%%%%%%%%%%%
%%%%%%%%%%% Introduction %%%%%%%%%%%
%%%%%%%%%%%%%%%%%%%%%%%%%%%%%%%%%%%%
\section{Introduction}
The Peccei-Quinn (PQ) mechanism\,\cite{Peccei:1977hh,Peccei:1977ur,Weinberg:1977ma,Wilczek:1977pj} 
is the most successful solution to the strong $CP$-problem.
There, the PQ-symmetry is assumed to be almost exact which is broken only by the axial anomaly of QCD. 
After its spontaneous breaking, the associated pseudo Nambu-Goldstone boson, the axion $a$, 
obtains a non-vanishing potential by non-perturbative effects of QCD through the axial anomaly. 
Eventually, the effective $\theta$-angle is dynamically tuned to be vanishing by the vacuum expectation value (VEV) of the axion. 

For a successful PQ-mechanism, however, it is required to circumvent a lot of constraints
put by extensive axion searches\,\cite[for review]{Agashe:2014kda}.
The most popular approach to evade those constraints is to make the axion couple to the Standard Model particles very feeble,
so that the axion is invisible~\cite{Kim:1979if,Shifman:1979if,Zhitnitsky:1980tq,Dine:1981rt}.
There, the decay constant of axion, $f_a$ (and hence the PQ-breaking scale),  is taken to be very large, {\it e.g.}, $f_a>10^9$\,GeV. 

Another approach to evade the constraints is to make the axion heavy (see {\it e.g.} \cite{Dimopoulos:1979pp,Tye:1981zy} for early attempts). 
Among various attempts, a successful idea was proposed in \cite{Rubakov:1997vp} where a mirror copy of the Standard Model was introduced.
By larger mass scales of the mirrored sector, the QCD axion is made heavy 
via the axial anomaly in the mirrored sector without spoiling the PQ solution to the strong $CP$-problem.
This idea has been incarnated by a model constructed in \cite{Fukuda:2015ana}
in which experimental, astrophysical and cosmological constraints are examined carefully
(see also \cite{Berezhiani:2000gh,Hook:2014cda,Albaid:2015axa,Barbieri:2016zxn} for relevant discussions).
Resultantly, it has been shown that the axion decay constant can be as low as $f_a\simeq {\cal O}(1)$\,TeV 
when the axion mass is rather heavy, $M_a>{\cal O}(0.1)\,{\rm GeV}$.
We call this model a visible heavy axion model.

One of the advantage of the heavy axion model with a moderate decay constant
is that  the model is durable against  explicit breaking of the PQ symmetry 
by  Planck suppressed operators which are generically expected in 
quantum gravity~\cite{Hawking:1987mz,Lavrelashvili:1987jg,Giddings:1988cx,Coleman:1988tj,Gilbert:1989nq,Banks:2010zn}.
For example, the shift in the effective $\theta$ angle is as small as of ${\cal O}(10^{-11})$ 
even in the presence of dimension five PQ-breaking operator for $f_a \simeq {\cal O}(1)$\,TeV
and $M_a = {\cal O}(1)$\,GeV.%
\footnote{See discussions in the appendix\,\ref{sec:breaking}.}

In this paper, we discuss dark matter candidates in the visible heavy QCD axion model.
In \cite{Fukuda:2015ana}, 
%it has been only discussed whether the mirrored  sector does not cause 
%cosmological problems and 
it has been deferred to discuss whether the mirrored sector provides good candidates for dark matter.
In fact, the model predicts multiple stable particles since the mirror copied sector possesses the same symmetry structure with the Standard Model sector.
They are the photon ($\gamma'$), the nucleons ($N'$), and two of the electron ($e'$), the lightest neutrino ($\nu'$) and the charged pion ($\pi'^\pm$)
in the mirrored sector.
Therefore, it is enticing to ask whether they can be good candidates for dark matter.

As we will show, $\pi'^\pm$ with masses in the TeV range can be a viable candidate for dark matter when it is lighter than all of $\nu'$.
It is also shown that $e'^\pm$ with a mass in the hundred GeV range can also be a viable candidate.
Notably, $\pi'^\pm$ and $e'^\pm$ serve as self-interacting dark matter with a long range force.
It should be noted that such darkly-charged dark matter is severely constrained~\cite{Ackerman:mha,Feng:2009mn,Feng:2009hw}.
Recently, however, it has been pointed out that there are a number of mitigating factors to the constraints,
which revives possibility of darkly-charged dark matter~\cite{Agrawal:2016quu}.
We also show that the mirrored neutron, $n'$, can be lighter than the mirrored proton, $p'$, and hence, be the lightest baryon in the mirrored sector.  
Accordingly,  it can also be  a viable dark matter candidate when its mass is around $100$\,TeV.
It is also shown that $\nu'$  also can be a viable candidate for dark matter.

The paper is organized as follows. 
In section~\ref{sec:model}, we briefly review the visible heavy axion model in \cite{Fukuda:2015ana}. 
In section~\ref{sec:darkmatter}, we discuss the dark matter candidates in the mirrored sector of the visible heavy axion model.
The final section is devoted to our conclusions and discussions.

%%%%%%%%%%%%%%%%%%%%%%%%%%%%
\section{Model of Visible Heavy QCD Axion}
\label{sec:model}
In this section, we first review a model of visible heavy QCD axion\,\cite{Fukuda:2015ana}. 
In this model, a 
copy of the standard model is introduced following the Rubakov's idea\,\cite{Rubakov:1997vp}. 
There, we assume  a $\mathbb Z_2$ exchanging symmetry between 
the Standard Model and its mirror copy. 
Due to the  $\mathbb Z_2$ symmetry,  the $\theta$-angles in these two sectors 
are aligned at the high energy input scale, {\it i.e.} $\theta = \theta'$. 
Throughout this paper, objects in the copied sector are referred with a prime ($^\prime$).

To implement the PQ mechanism, we introduce QCD colored left-handed Weyl fermions, $\psi_L$ and $\bar\psi_R$, 
and those for QCD${}'$, $\psi'_L$ and $\bar\psi'_R$.
We choose the PQ charges of $\psi_L$ and $\psi'_L$ to be $0$ and the ones of $\bar \psi_R$ and $\bar\psi'_R$ to be $-1$.
A complex scalar $\phi$ with a PQ charge $+1$  is introduced to break the PQ-symmetry spontaneously.
As in the KSVZ axion model~\cite{Kim:1979if,Shifman:1979if}, $\phi$ couples to  $\psi$ and $\psi'$ via
\begin{eqnarray}
\label{eq:PQInt}
\Delta \mathcal L = g\phi \psi_L\bar{\psi}_R + g\phi \psi'_L\bar{\psi}'_R + \text{H.c.},
\end{eqnarray}
where $g$ is a coupling constant.
Here, we assume that $\phi$ is even under the  $\mathbb Z_2$  symmetry.

Assuming that $\phi$ obtains a VEV, we decompose $\phi$ into an axion $a$ and a scalar boson $s$,
\begin{eqnarray}
\phi = \frac{1}{\sqrt{2}}(f_a + s) e^{i a/f_a}\ .
\label{eq:phi}
\end{eqnarray}
Here, $f_a$ is the decay constant of the axion.
Due to the VEV of $\phi$, $\psi$'s become heavy vector-like quarks with masses
\begin{eqnarray}
\label{eq:massextra}
m_{\psi}^{(\prime)} = \frac{g}{\sqrt{2}} f_a\ .
\end{eqnarray}
We additionally introduce small mixings between $d^{(\prime)}_i$ ($i=1,2,3$) quarks in the Standard Model${}^{(\prime)}$ and $\psi^{(\prime)}$ by assuming appropriate gauge charges,
\begin{eqnarray}
\label{eq:SMMixing}
\Delta \mathcal L = \varepsilon_i \mu \psi_L\bar d_{Ri} + \varepsilon_i \mu \psi'_L\bar d'_{Ri}\ ,
\end{eqnarray}
where $\varepsilon_i \ll 1$ are small mixing parameter and the $\mu$ is a representative mass scale of ${\cal O}(m_\psi)$.
Through the mixing term, $\psi$'s decay into Standard Model and corresponding mirror sector quarks.%
\footnote{We may instead assume mixings between $u^{(\prime)}$ quarks and $\psi^{(\prime)}$.}

It should be emphasized that the axion is common among the Standard Model and its mirrored copy.
With a single axion, the effective $\theta$ angles of QCD and QCD${}'$ are simultaneously set to be zero
due to the $\mathbb Z_2$ symmetry. Since $\theta$ and $\theta'$ hardly run under the renormalization group evolution\,\cite{Ellis:1978hq}, 
they are aligned even below the spontaneous breakdown of the $\mathbb Z_2$ symmetry.
Because of the breaking, the dynamical scale of QCD$'$ can become much higher than that of QCD (see  \cite{Fukuda:2015ana} for details).
With a large dynamical scale of QCD$'$, the axion obtains the mass dominantly from QCD$'$,
\begin{eqnarray}
\label{eq:AxionMass}
{M_a}^2 \simeq \frac{m_u'm_d'}{(m_u' + m_d')^2}\frac{m_{\pi}'^2f_\pi'^2}{{f_a}^2},
\end{eqnarray}
where $m_{u,d}'$ are the masses of $u'$ and $d'$ quarks, $m_\pi'$ the mass of $\pi'$,
$f_\pi'$ the decay constant of $\pi'$.
In terms of the dynamical scale of QCD$'$ and the VEV of Higgs$'$, $v_{EW}'$,
those quantities are given by
\begin{eqnarray}
\label{eq:massprime}
m_{u,d}' \simeq m_{u,d} \times \frac{v_{EW}'}{v_{EW}}, \quad
m_{\pi}'^2 \simeq m_{\pi}^2 \times \frac{\Lambda_{\rm QCD}'}{\Lambda_{\rm QCD}} \frac{v_{EW}'} {v_{EW}} \ ,
\quad
f_{\pi}' \simeq f_{\pi} \times \frac{\Lambda_{\rm QCD}'}{\Lambda_{\rm QCD}} \ .
\end{eqnarray}
In the following analysis, we assume that  the $\mathbb Z_2$ exchanging symmetry is softly (or spontaneously) broken
and take $\Lambda_{\rm QCD}' $ and $v_{EW}'$ are independent parameters (see \cite{Fukuda:2015ana}
for concrete examples). Note that if $\Lambda'_{\rm QCD}$ is greater than the tree-level Higgs$'$ VEV, 
the electroweak symmetry is broken by $\Lambda'_{\rm QCD}$ and $v'_{\rm EW}\sim\Lambda'_{\rm QCD}$ is induced.

%%%%%%%%%%%%%%%%%%%%%%%%%%%%
\begin{figure}[t]
\begin{center}
\begin{minipage}{.46\linewidth}
  \includegraphics[width=\linewidth]{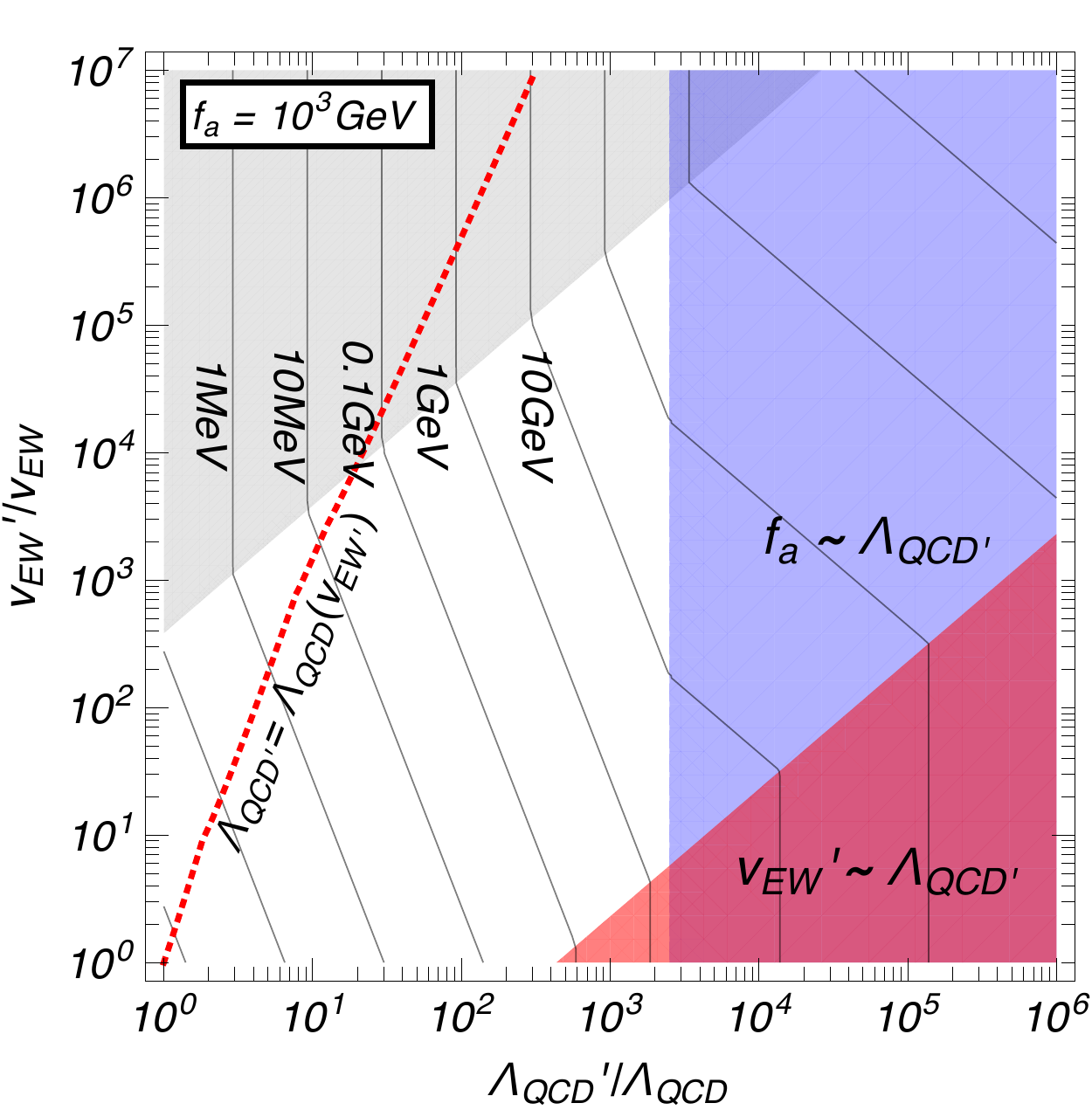}
 \end{minipage}
 \hspace{1cm}
 \begin{minipage}{.46\linewidth}
  \includegraphics[width=\linewidth]{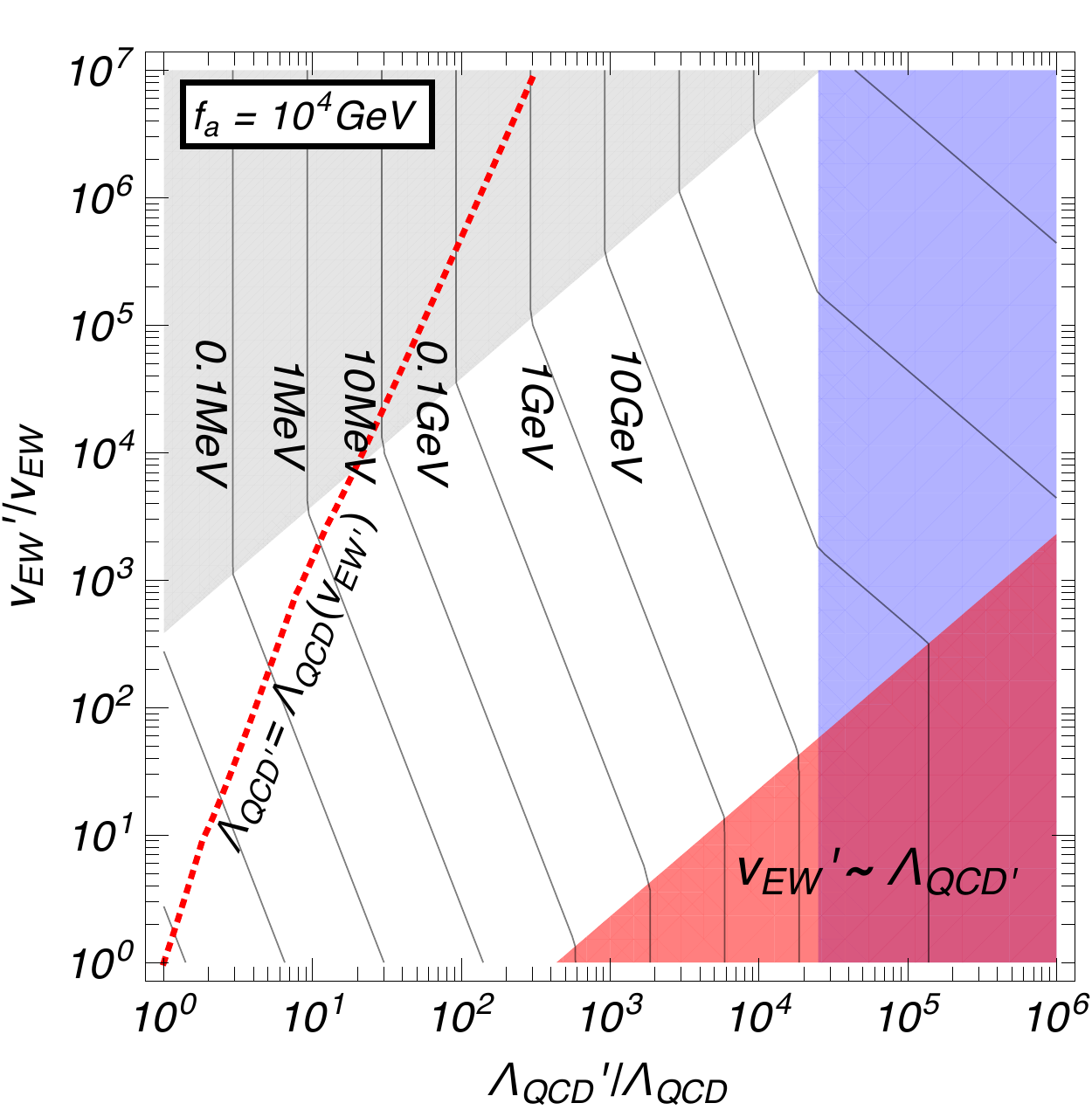}
 \end{minipage}
 \end{center}
\caption{\sl \small
Contour plots of the axion mass as a function of $\Lambda_{\rm QCD}'$ and $v_{\rm EW}'$ for given $f_a$.
Here, we take $f_\pi \simeq 93$\,MeV, $m_u'/m_d' \simeq m_u/m_d = 0.56$,  $\Lambda_{\rm QCD} \simeq 400$\,MeV and $v_{EW} \simeq 174$\,GeV. 
In the gray shaded regions, the quark masses in the mirrored sector are larger than $\Lambda_{\rm QCD}'$
where the axion mass does not depend on the quark mass any more.
In the blue shaded regions, the PQ-symmetry breaking is caused by the condensation of $\psi_L'\bar\psi_R'$ due to
the strong dynamics and hence $f_a = {\cal O}(\Lambda_{\rm QCD}')$.
In the red shaded regions, the electroweak symmetry breaking and the VEV of Higgs$'$ in the mirrored sector is caused by
the condensations of quarks$'$, leading to $v_{\rm EW}' \simeq {\cal O}(\Lambda_{\rm QCD}')$.
}
\label{fig:ma}
\end{figure}
%%%%%%%%%%%%%%%%%%%%%%%%%%%

In Fig.\,\ref{fig:ma}, we show  contour plots of the axion mass as a function of $\Lambda_{\rm QCD}'$ and $v_{\rm EW}'$ for given $f_a$.
Here, we take $f_\pi \simeq 93$\,MeV, $m_u'/m_d' \simeq m_u/m_d = 0.56$,  $\Lambda_{\rm QCD} \simeq 400$\,MeV and $v_{EW} \simeq 174$\,GeV. 
In the gray shaded regions, the quark masses in the mirrored sector are larger than $\Lambda_{\rm QCD}'$
where the axion mass does not depend on the quark mass any more.%
\footnote{In the figure, the boundary between these two regimes is taken to
$m_{\pi}'$ in Eq.\,(\ref{eq:massprime}) is equal to $m_u'+m_d'$.}
We call this region as the heavy quark region.
In the blue shaded regions, the PQ-symmetry breaking is caused by the condensation of $\psi_L'\bar\psi_R'$ due to
the strong dynamics and hence $f_a = {\cal O}(\Lambda_{\rm QCD}')$.
In the red shaded regions, the electroweak symmetry breaking in the mirrored sector 
is caused by the condensations of quarks$'$, leading to $v_{\rm EW}' \simeq {\cal O}(\Lambda_{\rm QCD}')$ as mentioned above.
In the figure, we also show the parameter region where $\Lambda_{\rm QCD}'$ is increased purely by the effects 
of larger quark masses in the mirrored sector due to a large $v_{\rm EW}'$ (red dashed lines).%
\footnote{The effects of  $\psi'$ contributions to the renormalization group running of the coupling constant 
of QCD$'$ do not cause visible difference in the figures even for $\Lambda_{\rm QCD}' \gg m_{\psi}'$.}

As discussed in \cite{Fukuda:2015ana}, the mirrored sector is in thermal equilibrium 
with the Standard Model sector in the early universe, via the axion exchange.
As the temperature of the universe decreases and becomes much lower than the axion mass, 
the mirrored sector decouples from the Standard Model sector.
Thus, when the axion is much heavier than the QCD phase transition temperature,  $T_{\rm QCD} = {\cal O}(100)$\,MeV, 
the contributions of the copied sector to the effective number of relativistic species are sufficiently suppressed 
due to $\Lambda_{\rm QCD}' \gg \Lambda_{\rm QCD}$.
For a lighter axion, on the other hand, $\gamma'$  decouples below $T_{\rm QCD}$ and contributes
the dark radiation, which causes tensions with the Big-Bang Nucleosynthesis and the Cosmic Microwave Background (CMB).
To avoid such problems, we concentrate on parameter regions where $M_{a} \gtrsim 1\,\text{GeV}$ in the following arguments.\footnote{As discussed in \cite{Chiang:2016eav}, the coupling between the axion and $\gamma'$ may be suppressed. In that case, the constraints on the axion mass come only from the following experiments.}

Let us also summarize the constraints and the visibility of the heavy axion model at collider experiments.
For a rather heavy axion, $M_a \gtrsim 3m_\pi$, the constraints from the beam dump experiments such as the CHARM experiment~\cite{Bergsma:1985qz} 
are not applicable due to its short lifetime.
The axion in this mass range is also free from the constraints from the rare $K$-meson decay~\cite{Artamonov:2009sz} since the axion mode is closed.
The constraints from the rare $B$-meson decays are also evaded due to the lack of the direct axion couplings to the quarks 
as in the case of the KSVZ axion model.

The LHC experiments put  lower limits on the mass of the extra quarks $\psi$.
The experimental lower limits on the extra quark masses 
are $800\mbox{--}900\,{\rm GeV}$~\cite{Aad:2014efa,Aad:2015tba,Khachatryan:2015gza,Khachatryan:2015oba}, 
depending on the branching ratios of $\psi$ into $b$ and $t$ quarks.
Assuming $g \sim 1$, the current constraints require $f_a\gtrsim1\,{\rm TeV}$.

The radial and the axion components of $\phi$ ({\it i.e.} $s$ and $a$) can be also produced at the LHC experiments
via the couplings to the gluons, when their masses are below a TeV range.
For example, the production cross sections of $s$ would be  ${\cal O}(100$--$1)\,{\rm fb}\times \left({1\,\text{TeV}}/{f_a}\right)^2$
for $M_s \simeq 500$\,GeV--$1$\,TeV, which mainly decays into a pair of the axions.
The majority of axions subsequently decay into a pair of jets for $M_a \gg {\cal O}(100)$\,MeV. A part of them decay into $2\gamma$, whose branching ratio is $\alpha^2/\alpha_s^2 \sim 0.01$ or more\,\cite{Chiang:2016eav}.
Since $s$ is much heavier than $a$, the final decay products of each axion are highly collimated and look like a single jet and photon, respectively. Comparing the branching ratio with the background, this one photon plus one jet channel may be most sensitive to search $s$.
For example, if we simply scale the current backgrounds at ATLAS $13$\,TeV search \,\cite{Aad:2015ywd}, we can conclude that it is possible to detect $s$ for the integrated luminosity $3$\,ab${}^{-1}$ in some parameter region. Once such an excess is observed, we can study the difference between a single photon and collimated photons\,\cite{Fukuda:2016qah}. Note that if $a\mbox{-}\gamma'\mbox{-}\gamma'$ coupling is suppressed, as is mentioned above, the axion may be as light as $3m_\pi\sim400\,$MeV. In that region, the branching ratios of $a\to2\gamma$ and $a\to\text{mesons}$ are comparable and diphoton like channel may be most sensitive\,\cite{Chiang:2016eav}.

%%%%%%%%%%%%%%%%%%%%%%%%%%%%%%%%%%%%%%%%%%%%%%%%%%%%%%
\section{Dark Matter Candidates in the Mirrored Sector}\label{sec:darkmatter}
\subsection{Stable Particles}
Stable particles in the mirrored sector are $\gamma'$, $N'$, and two of $\nu'$, $e'$ and $\pi'^{\pm}$.
In the minimal model of the visible heavy axion model, each the Standard Model and the mirrored sector
has a single Higgs doublet, and hence,  $\text{U}(1)_{\rm QED}$  and $\text{U}(1)_{\rm QED}'$ 
are not broken spontaneously.
Thus, $\gamma'$ is massless and stable.
The stabilities of other particles are associated with symmetries, {\it i.e.} $B'$, $L'$ and $Q_{\rm QED}'$ symmetries.

In the Standard Model sector, we assume the seesaw mechanism to account for the tiny neutrino masses~\cite{Yanagida:1979as,Ramond:1979py}~\cite[see also][]{Minkowski:1977sc}.
If the seesaw mechanism also works in the mirrored sector, the neutrino masses in the mirrored sector, $m_\nu'$, is enhanced by $(v_{EW}'/v_{EW})^2$,
which easily exceeds the upper limit on the hot dark matter mass,  $m_\nu' \ll {\cal O}(10)$\,eV, 
from CMB lensing  and cosmic shear~\cite{Osato:2016ixc}.%
\footnote{Here, we roughly translate the constraint on the gravitino mass, $m_{3/2} \lesssim 4.7$\,eV (95\%C.L.)~\cite{Osato:2016ixc},
by assuming that the decoupling temperature of $\nu'$ from the thermal bath of the Standard Model sector is similar to the gravitino.}
To evade this constraint, we assume that the seesaw mechanism does not take place in the mirrored sector. 
This can be achieved by turning off spontaneous breaking of the $B'-L'$ symmetry in the mirrored sector 
so that  the Majorana masses of the right-handed neutrinos in the mirrored sector vanish (see \cite{Fukuda:2015ana}  for details).

When the spontaneous breaking of the $B'-L'$ symmetry is turned off,
thermal leptogenesis~\cite{Fukugita:1986hr}~\cite[see][for review]{Giudice:2003jh,Buchmuller:2005eh,Davidson:2008bu}
does not take place in the mirrored sector. 
Accordingly, there is no $B'$ asymmetry in the mirrored sector when the $B$ asymmetry in the Standard Model sector is provided by thermal leptogenesis.
This feature is important  for the $N'$ relic density not to exceed the observed dark matter density even for $m_{N}' \gg 1$\,GeV.

In this set up, $\nu'$s obtain the Dirac neutrino masses via the Yukawa interaction 
to the Higgs boson.
Depending on the Yukawa coupling, $\nu'$s can be lighter or heavier than $\pi^{\prime\pm}$.
When (at least one of) $\nu'$s are lighter than $\pi'^\pm$, 
$\pi'^\pm$ decays into a pair of charged lepton$'$ and $\nu'$. 
On the other hand, $\pi'^\pm$ becomes 
stable when all the $\nu'$s are heavier than $\pi'^\pm$.
Therefore, the stable particles in the mirrored sector are
\begin{eqnarray}
\left\{
\begin{array}{lllll}
\gamma'\ , &  e' \ , & \pi^{\prime\pm} \ , & N' \ , & ({\mbox{for }}m_\nu' > m_{\pi^\pm}' )\ , \\
\gamma'\ , & e' \ , & \nu' \ , & N'  \ , &  ({\mbox{for }}m_\nu' < m_{\pi^\pm}') \ .
\end{array}
\right.
\end{eqnarray}
In the following, we discuss whether we have good dark matter candidates in each possibility.

Let us comment here that  $m_\nu' \ll m_{\pi^\pm}'$ can be automatically achieved if there are only two generations of the right-handed neutrinos in each sector. 
In fact, the lightest $\nu$ and $\nu'$ are both massless.
It should be also noted that two generations of the right-handed neutrinos are enough for successful thermal leptogenesis 
in the Standard Model sector\,\cite{Frampton:2002qc,Raidal:2002xf,Ibarra:2003up,Harigaya:2012bw}. 

%%%%%%%%%%%%%%%%%%%%%%%%
\subsection{Masses of Dark Matter Candidates}
\begin{figure}[t]
\begin{center}
\begin{minipage}{.325\linewidth}
{\sl \small $N'$}  
\includegraphics[width=\linewidth]{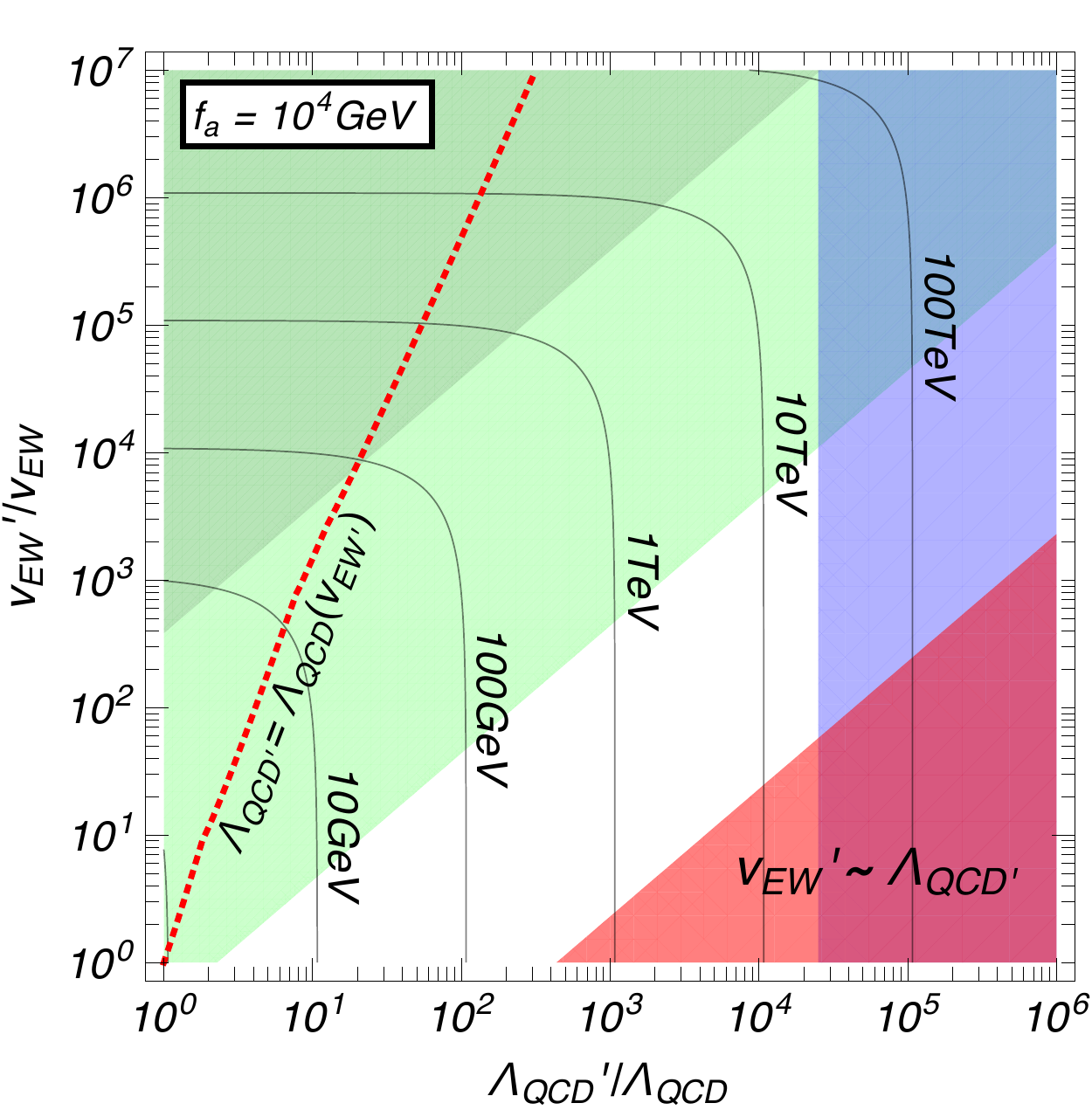}
 \end{minipage}
 \begin{minipage}{.325\linewidth}
 {\sl \small $\pi^{\prime\pm}$}  
  \includegraphics[width=\linewidth]{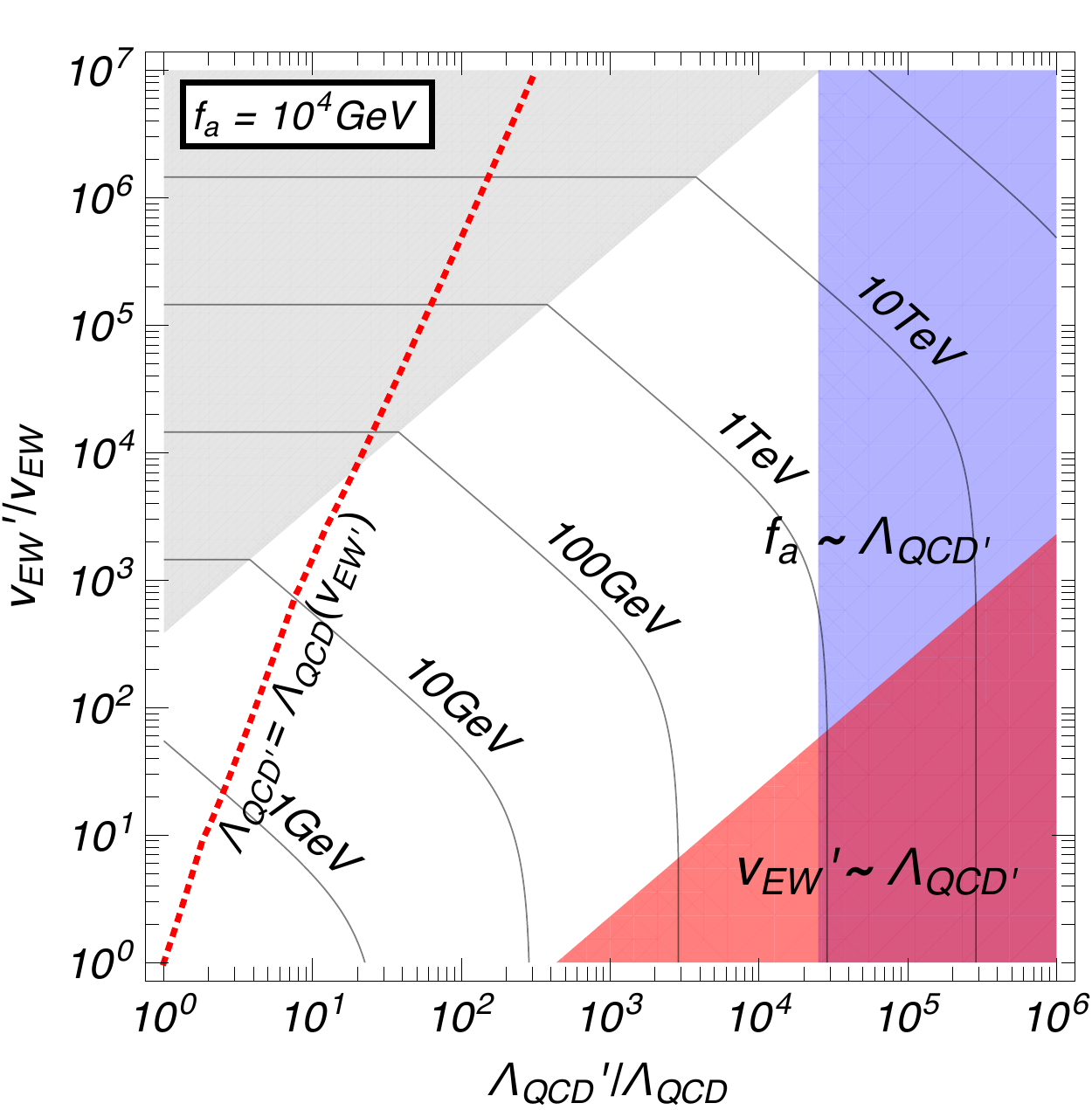}
 \end{minipage}
  \begin{minipage}{.325\linewidth}
    {\sl \small $e^{\prime}$}  
  \includegraphics[width=\linewidth]{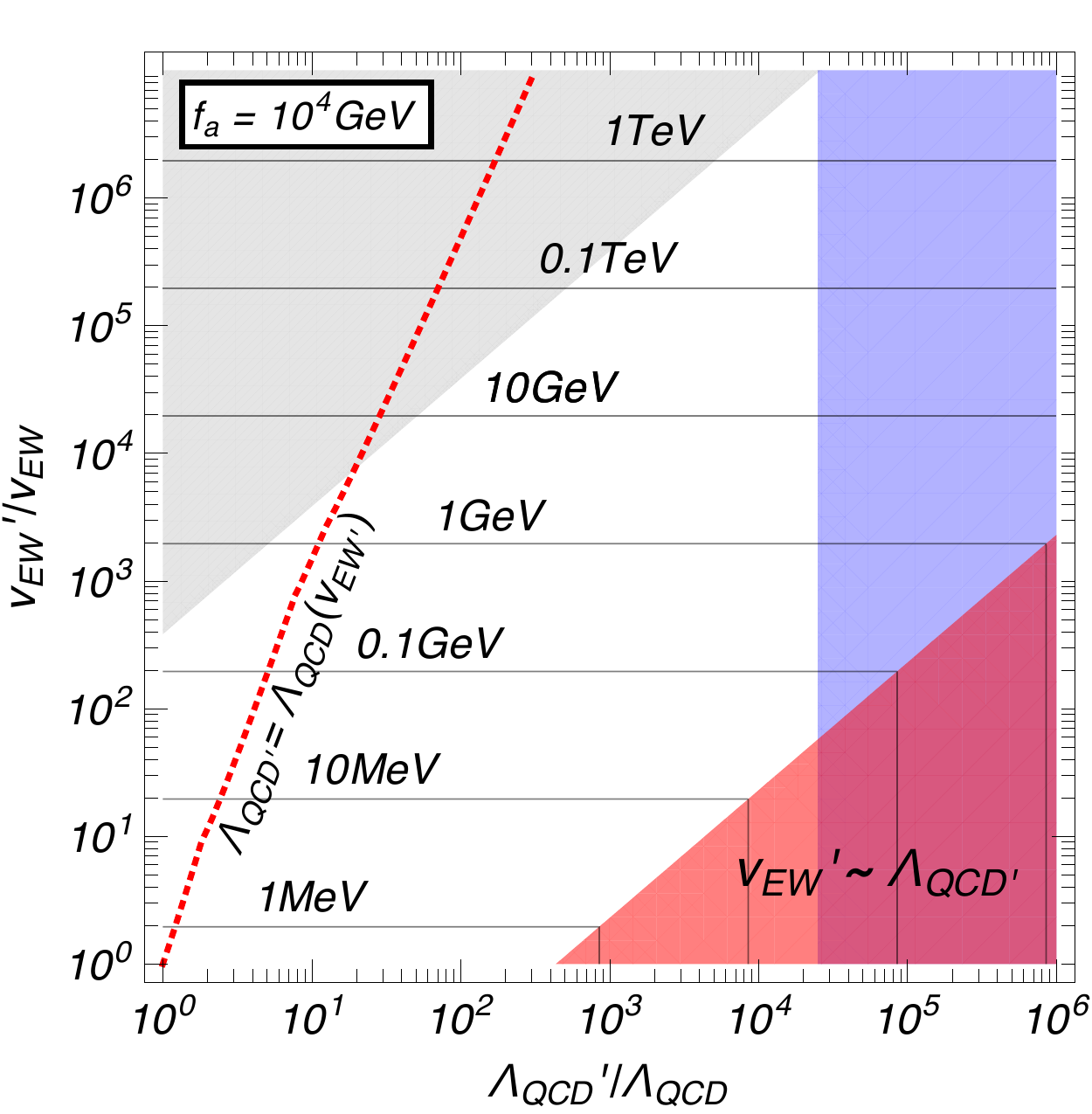}
 \end{minipage}
 \end{center}
\caption{\sl \small
Contour plots of the masses of $N'$, $\pi'^\pm$ and $e'$.
The gray, blue and red shaded regions are the same with the ones in Fig.\,\ref{fig:ma}.
The axion decay constant $f_a$ does not affect the masses of the stable particles,
 which only shifts the blue shaded regions. 
 In the green shaded region of the mass of $N'$, $p'$ is lighter than
 $n'$, while $n'$ is lighter in the other region.
}
\label{fig:masses}
\end{figure}

In Fig.\,\ref{fig:masses}, we show the masses of the stable particles.
The average nucleon mass is approximately estimated by
\begin{eqnarray}
m_{N'} \equiv \frac{m_n' + m_p'}{2} &\simeq& \left(\frac{m_n + m_p}{2} - 3 \bar{m}\right) \times \frac{\Lambda_{\rm QCD}'}{\Lambda_{\rm QCD}}  + 3 \bar{m} \times \frac{v_{EW}'} {v_{EW}} \ , 
\end{eqnarray}
where $\bar{m}$ is an average of the $u$ and $d$ quark masses,  $m_u = 2.2^{+0.6}_{-0.4}$\,MeV
and $m_d = 4.7^{+0.5}_{-0.4}$\,MeV~\cite{Olive:2016xmw}.%
\footnote{There is an ${\cal O}(1)$ ambiguity for the quark$'$ mass contributions for $\Lambda_{\rm QCD}' \ll \bar{m}'$.
However, the contributions from the quark$'$ mass to $m_N'$ is only important when the quark$'$ mass
is larger than $\Lambda_{\rm QCD}'$, where nucleon mass can be approximated by $3\times\bar{m}'$.}
The $N'$ masses are dominated by the masses of the quark$'$ 
when the quark$'$ masses are heavier than $\Lambda_{\rm QCD}'$.%
\footnote{For $v_{EW}' \gg 10^{5-6}\times v_{EW}$,  $m_\psi'$ can be smaller than $m_{u,d}'$ for $f_a \simeq 10^3$\,GeV. 
In such region, the lightest baryon consists of $\psi'$s, and hence, 
the $N'$ mass in the figure for $v_{EW}' \gg 10^{5-6}\times v_{EW}$ should not  be taken literally.
}

The mass difference between the neutron$'$ ($n'$)  and the proton$'$ ($p'$) is estimated by
\begin{eqnarray}
m_{n'} - m_{p'} &\simeq& \delta m_{n-p}^{\rm QED}  \times \frac{\Lambda_{\rm QCD}'}{\Lambda_{\rm QCD}} 
+ \kappa_N (m_d-m_u)\times \frac{v_{EW}'} {v_{EW}}\ , 
\end{eqnarray}
where $ \delta m_{n-p}^{\rm QED}$ denotes the electromagnetic contribution to the $n$--$p$ mass difference,
and $\kappa_N$  parameterizes the isospin-violating contribution.
As leading order approximations, we use the central values of the Standard Model~\cite{Walker-Loud:2014iea}
\begin{eqnarray}
\delta m_{n-p}^{\rm QED} &=& - 0.178^{+0.0004}_{-0.064}\,{\rm GeV}  \times \alpha_{\rm QED} \ ,\\
\k_N &=& 0.95^{+0.08}_{-0.06} \ .
\end{eqnarray}
Remarkably, $n'$ can be lighter than  $p'$ when $\Lambda_{\rm QCD}'$ becomes very large.
In fact, in the green shaded region in Fig.\,\ref{fig:masses}, $p'$ is lighter than $n'$, while $n'$ is lighter in the other region.
It should be also noted that the mass difference is smaller than $m_{\pi^\pm}'$ in the entire parameter region,
and hence, both of $p'$ and  $n'$ are stable for $m_\nu' > m_{\pi^\pm}'$. 
If one of the neutrino$'$ mass and $m_e'$ is light enough, on the other hand, the heavier $N'$  
can decay into the lighter one.

The mass of $\pi'^0$ is estimated to be
\begin{eqnarray}
\label{eq:pi0}
m_{\pi^0}'^2 \simeq m_{\pi^0}^2 \times \frac{\Lambda_{\rm QCD}'}{\Lambda_{\rm QCD}} \frac{v_{EW}'} {v_{EW}} 
\end{eqnarray}
for $m_u'+m_d'  < m_{\pi^0}'$.
For $m_u', m_d' \gtrsim \Lambda'$, 
It is dominated by $m_u' + m_d'$ in the heavy quark mass region.%
\footnote{In the parameter region where $m_\psi'$ is smaller than $m_{u,d}'$, the lightest meson 
consist of $\psi'$. Thus, again, the mass of the pion in the figure should not be taken literally.}
The mass of $\pi^\pm$ is, on the other hand, given by,
\begin{eqnarray}
\label{eq:pi0}
m_{\pi^\pm}'^2 \simeq m_{\pi^0}'^2 + \alpha_{ QED}' \Lambda_{ QCD}'^2\ ,
\end{eqnarray}
where $\alpha_{\rm QED}'$  is the fine-structure constant of the QED$'$.

Finally, the mass of $e'$ is given by,
\begin{eqnarray}
m_e' = m_e \times  \frac{v_{EW}'} {v_{EW}}\ .
\end{eqnarray}
It should be noted that the $\mu'$ decays into $3e'$ via box diagrams in which $W$' boson circulate.
Thus, $\mu'$ cannot be a candidate for dark matter.

%%%%%%%%%%%%%%%%%%%%%%%%%%%
%%%%%%%%%%%%%%%%%%%%%%%%%%%

\subsection{Dark Matter Candidates For $m_\nu' > m_{\pi^\pm}'$}
\begin{figure}[t]
\begin{center}
\begin{minipage}{.325\linewidth}
{\sl \small $N'$}  
\includegraphics[width=\linewidth]{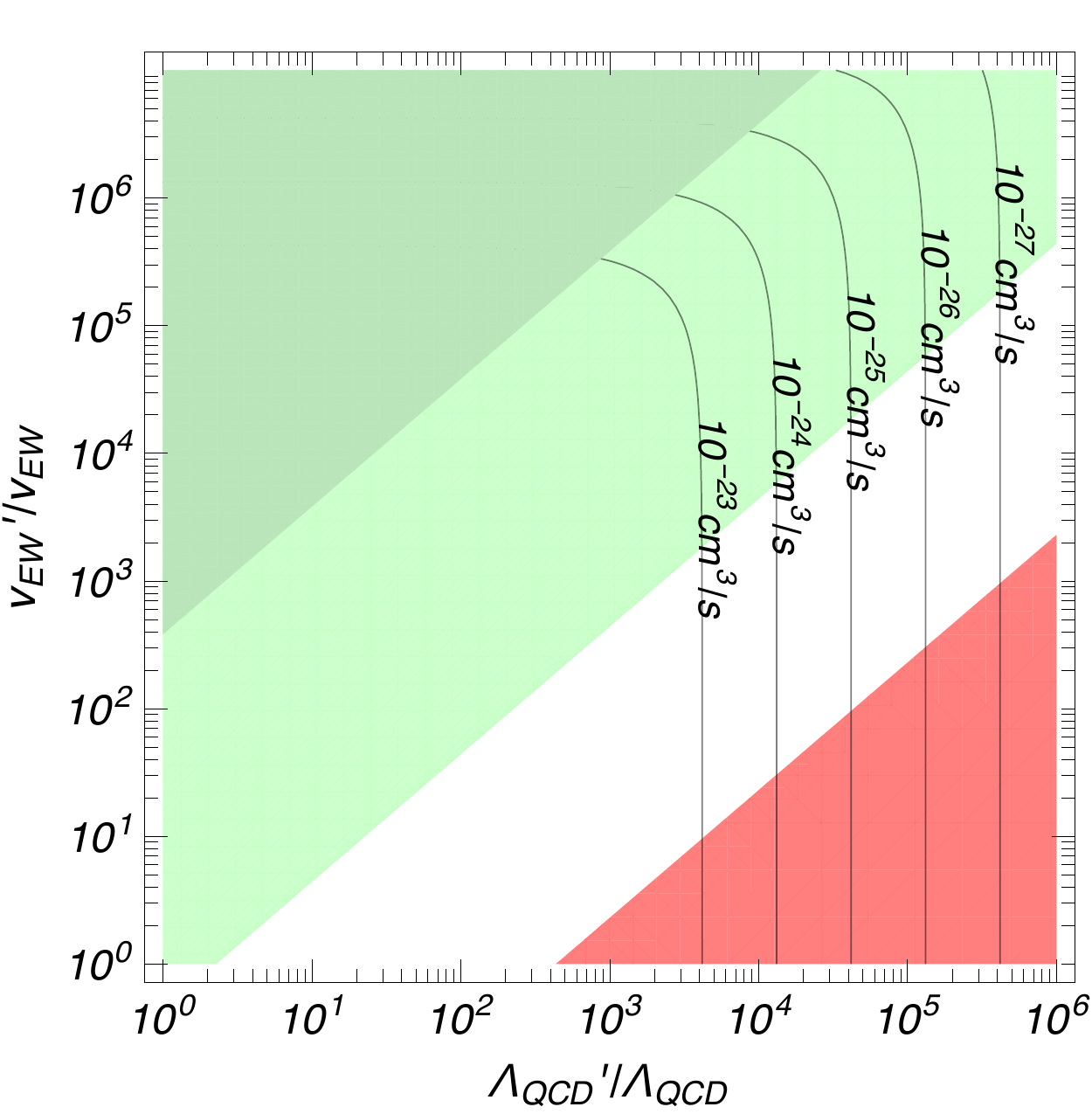}
 \end{minipage}
 \begin{minipage}{.325\linewidth}
 {\sl \small $\pi^{\prime\pm}$}  
  \includegraphics[width=\linewidth]{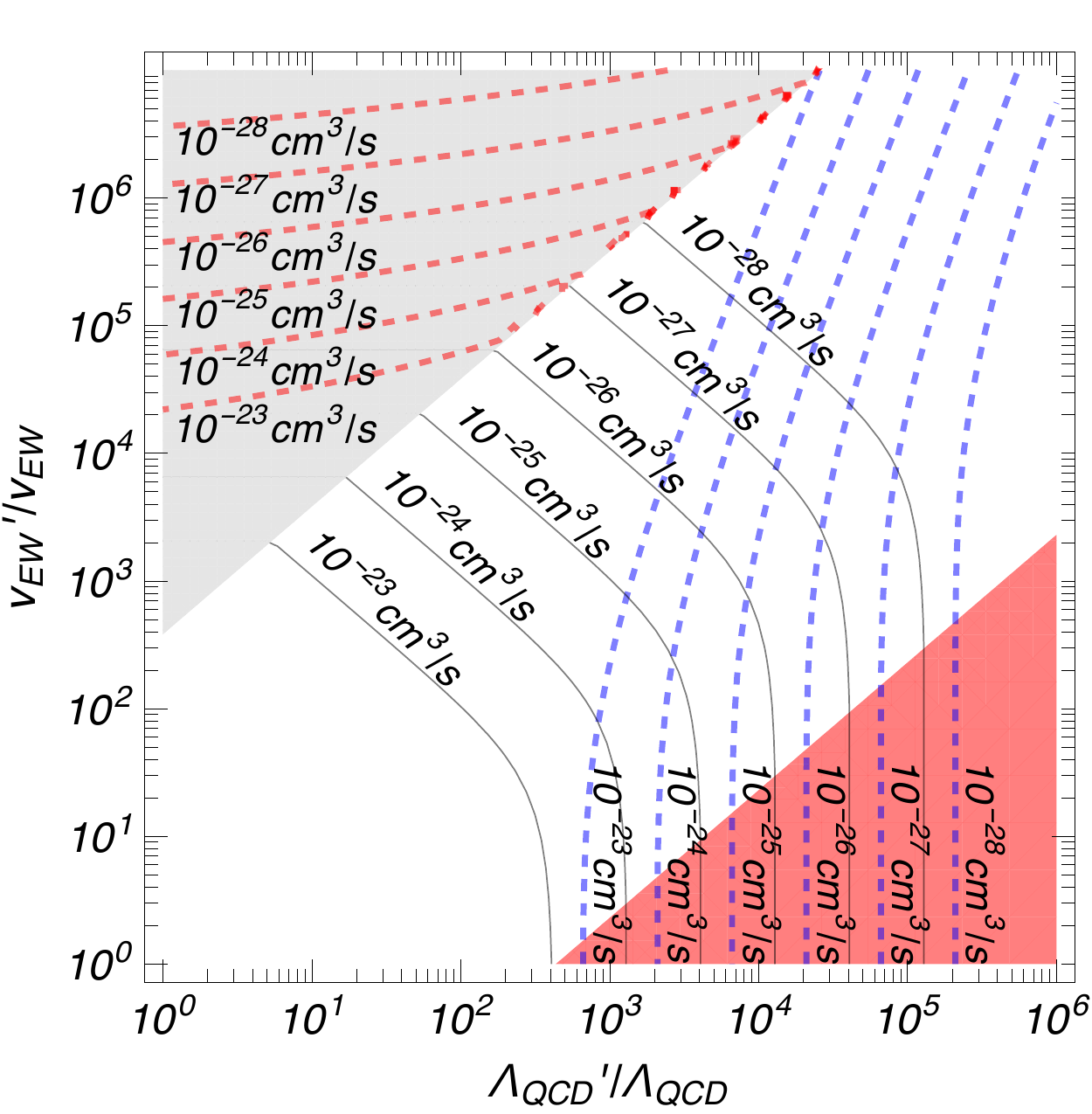}
 \end{minipage}
  \begin{minipage}{.325\linewidth}
    {\sl \small $e^{\prime}$}  
  \includegraphics[width=\linewidth]{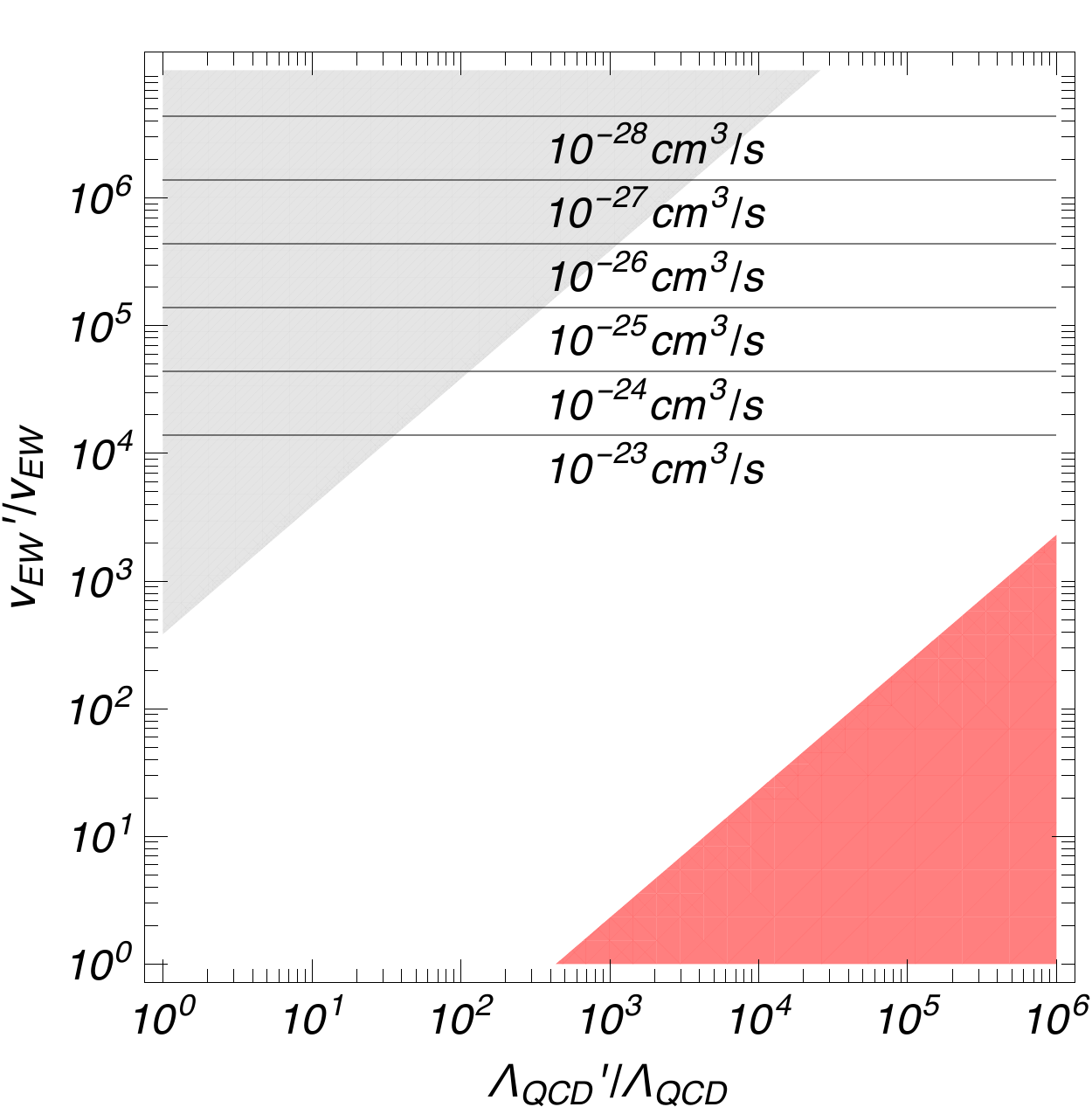}
 \end{minipage}
 \end{center}
\caption{\sl \small
Contour plots of the annihilation cross sections of $N'$, $\pi'^\pm$ and $e'$.
The gray, red, and green  shaded regions are the same with the ones in Fig.\,\ref{fig:masses}.
In the left panel, $N'$s annihilate into a pair of $\pi'$s.
In the central panel, $\pi^\pm$ annihilates into a pair of $\gamma'$ (solid) and 
into a pair of $\pi^0$ (dashed) in the region of $m_{\pi^\pm}' > m_u' + m_d'$.
In the heavy quark$'$ region, we show the annihilation cross section of $d'$ into 
a pair of gluon$'$s.
In the right panel, $e'$s annihilate into a pair of $\gamma'$s.
}
\label{fig:CS}
\end{figure}

First, let us discuss dark matter candidates for $m_\nu' > m_{\pi^\pm}$,
where $n'$, $p'$, $\pi'^\pm$ and $e'$ are stable.
To explain the observed dark matter density, 
$\Omega h^2 \simeq 0.1198\pm 0.0015$~\cite{Ade:2015lrj},
the averaged annihilation cross section of dark matter should be of 
\begin{eqnarray}
\label{eq:WIMP}
\vev{\sigma v} \sim 3\times 10^{-26} {\rm cm}^3/{\rm s}\ ,
\end{eqnarray}
\cite{Gondolo:1990dk} (see also \cite{Steigman:2012nb}.)
In Fig.\,\ref{fig:CS}, we show the annihilation cross sections of $N'$, $\pi'^\pm$, and $e'$
as functions of $\Lambda_{\rm QCD}'$ and $v_{\rm EW}'$.

In the figure, we assume that the annihilation cross section of $N'$ into $\pi'$s
saturates the so-called unitarity limit~\cite{Griest:1989wd},
\begin{eqnarray}
\vev{\sigma v_{\rm rel}} \sim \frac{8\pi}{m_{N}'^2} \ ,
\end{eqnarray}
where we approximate $v_{\rm rel}^2 \simeq 1/4$.
From the left panel of Fig.\,\ref{fig:CS}, we find that  $N'$ provides the observed dark matter 
density for $m_{N'} \sim 100$\,TeV if they are the sole dark matter candidate.

%%%%%%%%%%%%%%%%%%%%%%%%%%%
%%%%%%%%%%%%%%%%%%%%%%%%%%%

In the central panel of the figure, we show the annihilation cross section
of $\pi'^\pm$ into a pair of $\gamma'$ and into a pair of $\pi'^0$.
The averaged annihilation cross section of $\pi'^\pm$ into $\gamma'$ is given by,
\begin{eqnarray}
\vev{\sigma v_{\rm rel}} = \frac{\pi \alpha_{\rm QED}'^2}{m_{\pi^\pm}'^2}\ .
\end{eqnarray}
The annihilation cross section into $\pi'^0$ is, on the other hand, given by 
\begin{eqnarray}
\label{eq:CSpion}
\vev{\sigma v_{\rm rel}} \simeq 
\frac{1}{16\pi} \frac{9}{4m_{\pi^\pm}'^2}
\frac{m_{\pi^\pm}'^4}{f_\pi'^4}
\frac{(s - 4 m_{\pi^0}'^2)^{1/2}}{2 m_{\pi^0}'}\ ,
\end{eqnarray}
where $s \simeq 4 m_{\pi^\pm}'^2 (1 + v_{\rm rel}^2/4)$ (see {\it e.g.}  \cite{Weinberg:1966kf}).
In the central panel of the figure, those cross sections are shown by the solid lines and the orange dashed lines, respectively.
The figure shows that the cross section of ${\cal O}(10^{-26})$\,cm$^3/$s is achieved for $m_{\pi^\pm}' \simeq 400$\,GeV 
when the mode into $\gamma'$'s is dominant  and $m_{\pi^\pm}' = {\cal O}(1)$\,TeV
when the mode into $\pi'^0$'s is dominant.

In the heavy quark$'$  region, we also show the annihilation cross section of $d'$ into gluon$'$'s,
\begin{eqnarray}
\label{eq:CSquark}
\vev{\sigma v_{\rm rel}} \simeq \frac{55} {216 }  \frac{\pi\alpha'^2_{\rm QCD}}{m_d'^2}\ .
\end{eqnarray}
Here, the fine structure constant of QCD$'$ is estimated by
\begin{eqnarray}
\alpha'_{\rm QCD} \simeq \left(\frac{11}{2\pi} \log \frac{m_d'}{\Lambda_{\rm QCD}'}\right)^{-1}\ .
\end{eqnarray}
The figure shows that the  cross section of ${\cal O}(10^{-26})$\,cm$^3/$s is obtained for 
$m_d' = {\cal O}(1)$\,TeV.
It should be noted that the cross sections in Eqs.\,(\ref{eq:CSpion}) and (\ref{eq:CSquark}) receive
large higher order corrections for $\Lambda_{\rm QCD}' \sim m_{u}' + m_d'$,
and hence, their values at  $\Lambda_{\rm QCD}' \sim m_{u}' + m_d'$ are not reliable.

Finally, we also show the annihilation cross section of $e'$ into a pair of $\gamma'$s.
The annihilation cross section of $e'$ into $\gamma'$ is given by,
\begin{eqnarray}
\vev{\sigma v_{\rm rel}} = \frac{\pi \alpha_{\rm QED}'^2}{2 m_{e}'^2}\ .
\end{eqnarray}
The  cross section of ${\cal O}(10^{-26})$\,cm$^3/$s is achieved for $m_e' \simeq 300$\,GeV.

\begin{figure}[t]
\begin{center}
 \begin{minipage}{.46\linewidth}
  \includegraphics[width=\linewidth]{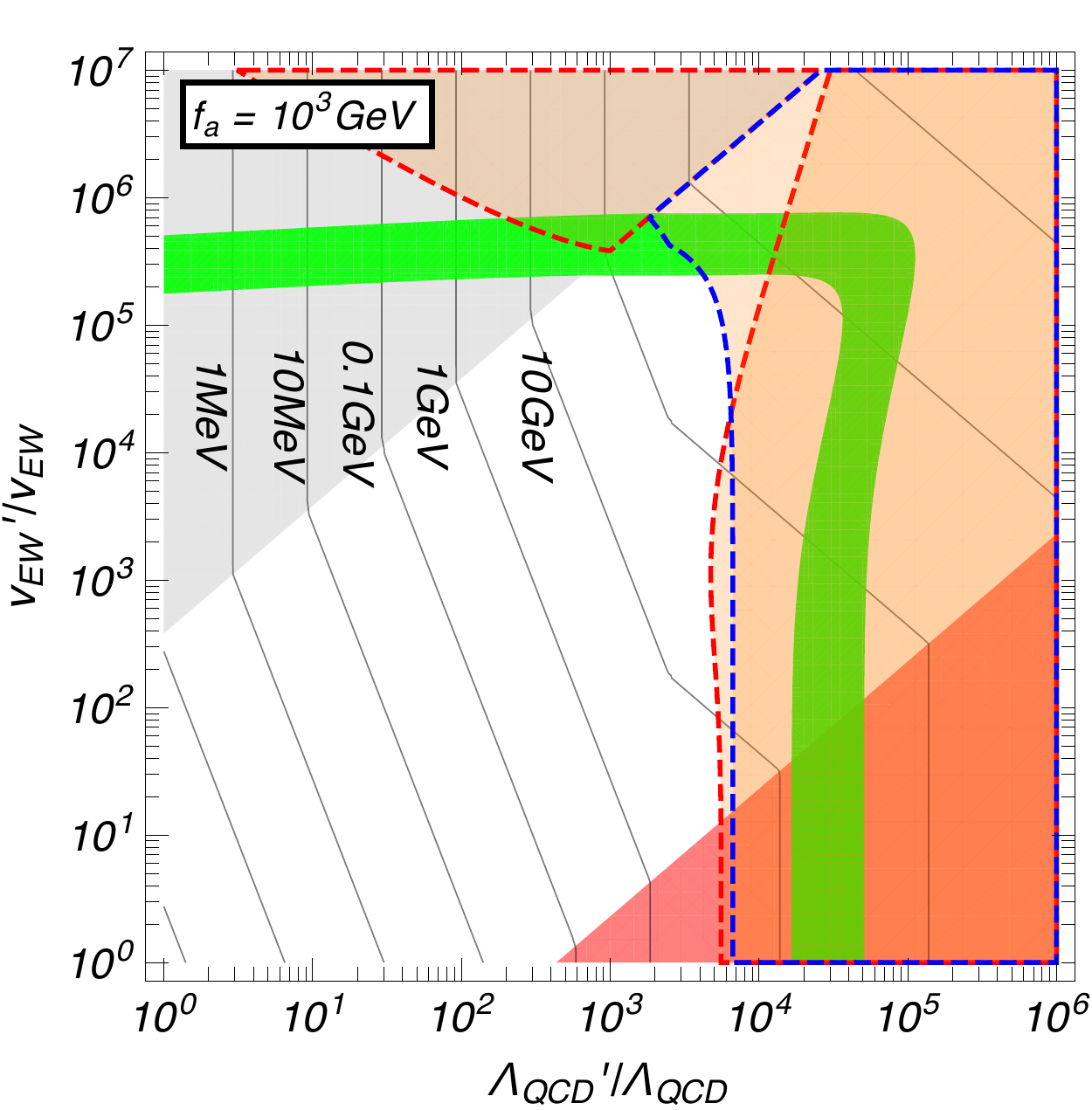}
 \end{minipage}
 \hspace{1cm}
 \begin{minipage}{.46\linewidth}
  \includegraphics[width=\linewidth]{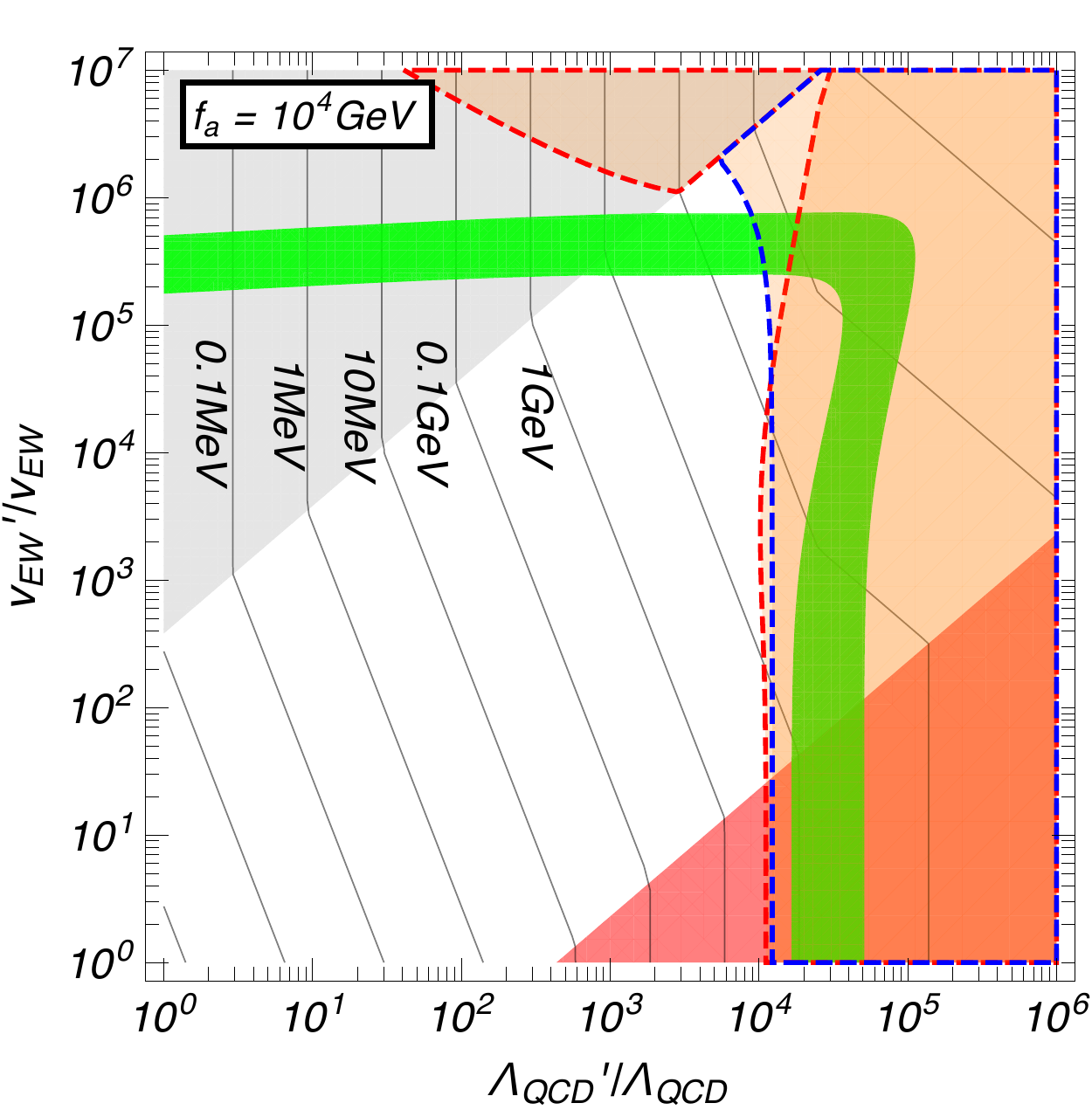}
 \end{minipage}

 \end{center}
\caption{\sl \small
The parameter region where the observed dark matter density is explained (green band)
for given  $f_a$ for $m_\nu' > m_{\pi^\pm}'$. There, the dominant components of the dark matter are ${\pi'^\pm}$, $e'$ and $e'$ and $\pi'^\pm$ for the vertical, curved and horizontal regions, respectively.
The contour plot of the axion mass is also shown.
The gray shaded regions are the same with the ones in Fig.\,\ref{fig:ma}.
The areas enclosed by the red and blue dashed lines are excluded by the 
the constraints on the dark matter annihilation from CMB observations.
The green band is not affected by $f_a$, while the CMB constraints 
get stringent for a smaller $f_a$.
}
\label{fig:relic1}
\end{figure}

Altogether, we show the parameter region where the observed dark matter density is explained in Fig.\,\ref{fig:relic1}
(green band).
To reflect our ignorance of the precise relation between 
the mass parameters ($\Lambda_{\rm QCD}'$, $v_{\rm EW}'$) with 
physical mass parameters and the interaction rates of hadron$'$, 
we show the parameter region where $\Omega h^2 = 0.03$--$0.3$ is achieved.
As the figure shows, the observed dark matter density can be explained 
by $\pi'^\pm$ with a mass in the TeV range for $m_{\pi^\pm}'>m_u' + m_d'$ ({\it i.e.} the vertical brach 
of the green band).
The dark matter density can be also explained by $e'$ with a mass around $300$\,GeV
for $\Lambda_{\rm QCD}'/\Lambda_{\rm QCD} \simeq 10^{3}$--$10^4$\,
GeV on the horizontal branch of the green band.
In the heavy quark$'$ region, dark matter consists of the mixture of the quark$'$ with a mass in the TeV range 
and $e'$ with a mass around $300$\,GeV.%
\footnote{The quark$'$ eventually confined into charged mesons. 
Here, we assume that the QCD$'$ dynamics which takes place after the dark matter freeze-out
does not affect the quark$'$ number density significantly (see {\it e.g.} discussions in \cite{Kang:2006yd,Harigaya:2016nlg} ).}
The relic density of $N'$ is subdominant in the favored region.

%%%%%%%%%%%%%%%%%%%%%%%%%%%%
%%%%%%%%%%%%%%%%%%%%%%%%%%%

It should be noted that dark matter components which annihilate into $\pi'^0$'s may lead the Standard Model jet via the $a$--$\pi'^0$ mixing with a mixing angle of ${\cal O}(f_\pi'/f_a)$.
 Furthermore, the annihilation cross section is significantly enhanced when the dark matter velocity becomes small
 since  $\pi'^\pm$ couples to the massless $\gamma'$.%
\footnote{For enhanced annihilation rate via the bound state formation, 
see \cite{Feng:2009mn,vonHarling:2014kha,Petraki:2015hla,An:2016gad,Petraki:2016cnz} }
The kinetic decoupling of darkly-charged dark matter takes place at around 
the temperature of the Standard Model sector to be,
\begin{eqnarray}
T_{\rm kd}\sim 0.5\,{\rm keV} \times \xi^{-7/3}\left(\frac{m_{DM}}{100\,\rm GeV}\right)^{5/3} \ ,
\end{eqnarray}
for $\alpha_{\rm QED}' \simeq 1/137$. 
Here, $\xi$ denotes the ratio between the temperatures of the mirrored sector and the Standard Model sector,
\begin{eqnarray}
\xi \equiv \frac{T_{\rm mirror}}{T} = \left(\frac{g_{*S}^{\rm mirror}(T_D)}{g_{*S}^{\rm mirror}(\xi T_{\rm kd}))}\right)^{1/3}
\left(\frac{g_{*S}(T_{\rm kd})}{g_{*S}(T_D)}\right)^{1/3}\ ,
\end{eqnarray}
with $g_{*S}$ and $g_{*S}^{\rm mirror}$ being the degrees of freedom of the Standard Model sector 
and the mirrored sector, respectively.
Thus, for example, the dark matter velocity at around the recombination time of the Standard Model sector is given by,
\begin{eqnarray}
v_{\rm DM} \sim 10^{-7}\times  \xi^{1/6} 
\left(\frac{100\,\rm GeV}{m_{DM}}\right)\ ,
\end{eqnarray}
with which the cross section is enhanced by the Sommerfeld enhancement factor,
\begin{eqnarray}
S \simeq  \frac{\pi \alpha_{\rm QED}'/v_{\rm DM}}
{1- e^{-\pi \alpha_{\rm QED}'/v_{\rm DM} }} \ .
\end{eqnarray}

It should be noted that the dark matter annihilation rate at around the recombination time is significantly constrained from CMB 
observations~\cite{Adams:1998nr,Chen:2003gz,Slatyer:2009yq,Kanzaki:2009hf,Galli:2009zc,Kawasaki:2015peu,Slatyer:2015jla,Cline:2013fm,Liu:2016cnk,Bringmann:2016din};
\begin{eqnarray}
\label{eq:CMB}
\frac{1}{2}\vev{\sigma v_{\rm rel}}  \lesssim  4 \times 10^{-25} \, {\rm cm}^3/{\rm s}\times \left(\frac{0.1}{f_{\rm eff}}\right)
\left(\frac{m_{DM}}{100\,\rm GeV}\right)\ ,
\end{eqnarray}
at 95\%C.L.~\cite{Ade:2015xua}.
Here, we use the efficiency factor $f_{\rm eff} \simeq 0.1$ which is the half of the one for the dark matter annihilation
into a pair of gluons~\cite{Slatyer:2015jla}.
In Fig.\,\ref{fig:relic1}, we show the parameter regions which are excluded by the CMB constraints on the annihilation 
cross section at around the recombination time.
Here, we scale the constraint in Eq.\,(\ref{eq:CMB}) by a factor of $(\Omega h^2/0.12)^2$ for each dark matter component.
The region enclosed by the red and blue dashed lines are excluded by the annihilation rate of $\pi'^\pm$ 
and $p'$ into the axion, respectively.
Here, we assume $\alpha_{\rm QED}' \simeq \alpha_{\rm QED}$.
The figure shows that the vertical branches of the green band where $\pi^\pm$ is the dominant dark matter component 
are excluded by the CMB observations.
It should be noted that $e'$  does not annihilate into the axion, and hence, the $e'$ component is not 
constrained by the CMB observations.

The dominant component of the dark matter discussed in this section are all charged under QED$'$, and hence, are self-interacting 
through a long-range force.
Such darkly-charged dark matter is severely constrained by the ellipticities of galaxy and cluster-scale dark matter halos, 
since the long-range interactions erase the non-sphericity~\cite{Ackerman:mha,Feng:2009mn,Feng:2009hw}. 
Among various constraints, the non-zero ellipticity of the gravitational potential of NGC720~\cite{Buote:2002wd} puts 
stringent constraints on the self-interaction cross section and excludes the darkly-charged
dark matter with $\alpha_{\rm QED}' \simeq 1/137$ for $m_{DM}\lesssim {\cal O}(1)$\,TeV~\cite{Feng:2009mn}.
Recently, however, it is pointed out that there are some uncertainties on the ellipticity of the inner parts of 
the galaxy and in the estimation of the timescale  to erase ellipticity, 
which revives the darkly-charged dark matter for $m_{DM}= {\cal O}(100)$\,GeV and $\alpha_{\rm QED}' =1/137$~\cite{Agrawal:2016quu}.
It is also pointed out out that there are  a number of mitigating factors
as for the constraints on the darkly-charged dark matter from the dwarf galaxy survival probability~\cite{Kahlhoefer:2013dca},
with which darkly-charged dark matter for $m_{DM}= {\cal O}(100)$\,GeV and $\alpha_{\rm QED}' \simeq1/137$ is consistent.
%The merging cluster constraints~%
%\cite{Markevitch:2003at,Randall:2007ph,2013ApJ...772..131D,Kahlhoefer:2013dca,Mohammed:2014iya,Dawson:2014jca,
%Harvey:2015hha,Massey:2015dkw,Kahlhoefer:2015vua,Randall:2016nxa,Robertson:2016xjh}
%\cite{Markevitch:2003at,Randall:2007ph,Peter:2012jh,Rocha:2012jg,Kahlhoefer:2013dca,
%Harvey:2015hha}
%on the long-range force between dark matter are also less stringent as noted in Ref.\,\cite{Agrawal:2016quu}.

Darkly-charged dark matter of $m_{DM} = {\cal O}(0.1$--$1)$\,TeV also has a huge self-interacting cross section per the dark matter mass 
of ${\cal O}(10^{2}\mbox{--}10^{4})$\,cm$^3/$s$/$g in dwarf galaxies for $\alpha_{\rm QED}' \simeq 1/137$~\cite{Agrawal:2016quu}. 
Such a large cross section affects the dark halo dynamics and could lead to core formation in dark halo~\cite{Kaplinghat:2015aga}. 
However, the effects of the huge self-interacting cross section per the dark matter mass 
of ${\cal O}(10^{2}\mbox{--}10^{4})$\,cm$^3/$s$/$g  require more detailed analysis as well as a larger statistical samples as noted in  \cite{Agrawal:2016quu}.
In view of these circumstances, we regard that darkly-charged dark matter candidates in this model are not ruled out currently 
and expect that future observations might be able to probe  intriguing features of the darkly-charged dark matter as self-interacting dark matter.

%%%%%%%%%%%%%%%%%%%%%%%%%%%%%%%%%%%%%%%%%%%%%%%%%%%%%%%%%%%%
\subsection{Dark Matter Candidates In the Presence of a Very Light $\nu'$}
Let us discuss next dark matter candidates when the lightest $\nu'$ is very light and stable.
Here, we require $m_{\nu'} \ll {\cal O}(10)$\,eV, so that  $\nu'$ evades the constraint from CMB lensing  and cosmic shear~\cite{Osato:2016ixc}.
As mentioned earlier, such a light $\nu'$ can be automatically achieved if there are only two generations of the right-handed neutrinos in each sector,
with which the lightest $\nu^{(\prime)}$ is massless in each sector.

In this case, $\pi'^\pm$ decays into $\nu'$, and hence, $\pi'^\pm$ is no more dark matter candidate.
Besides, the mass difference between $p'$ and $n'$ 
is larger than $m_e'$ in most parameter region, and hence, $n'$ decays into $p'$ 
for $m_n' > m_p'$ (i.e. in the green shaded region in Fig.\,\ref{fig:masses})
while $p'$ decays into $n'$ for $m_n' < m_p'$.
In the heavy quark$'$ mass region, on the other hand, the lightest and stable baryon corresponds to $\Delta'^{++} (u'u'u')$ baryon.%
\footnote{Here, we assume that $\psi'$ mixes with $d'$ as in Eq.\,(\ref{eq:SMMixing}) and $\psi'$ is heavier than $u'$ and $d'$, so that both $d'$ and $\psi'$ decay.}
As a result, the dark matter candidates in the presence of a very light (or massless) $\nu'$ are
\begin{eqnarray}
\left\{
\begin{array}{lll}
e' \ ,  & n' \ ,&  (m_p' > m_n' + m_e' )\ , \\
 e' \ , & p'  \ , &  (m_n' > m_p' + m_e') \ , \\
 e' \ , & \Delta'^{++}\ ,  &  (\mbox{in the heavy quark$'$ region}) \ . \\
\end{array}
\right.
\end{eqnarray}

It should be noted that the very light (or massless) $\nu'$ does not give a  visible contribution to the 
the effective number of relativistic species, $N_{\rm eff}$, as long as $T_D \gg T_{\rm QCD}$.
In fact, $N_{\rm eff}$  deviates from the Standard Model prediction, $N_{\rm eff}^{\rm SM} = 3.046$~\cite{Mangano:2005cc}
by
\begin{eqnarray}
{\mit \D} N_{\rm eff} = 
\left(
2 \left(\frac{11}{4}\right)^{4/3}
+ \frac{7}{8}\times 4
\right)\left(\frac{2}{g_{*S}(T_{D})}\right)^{4/3}
\times \left(\frac{7}{4} \left(\frac{4}{11}\right)^{4/3}
\right)^{-1} \simeq 0.18\ ,
\end{eqnarray}
which is consistent with the $N_{\rm eff}$ obtained from the CMB observation,
$N_{\rm eff} = 3.15\pm 0.23 $\,(68\,\%C.L.).

\begin{figure}[t]
\begin{center}
 \begin{minipage}{.46\linewidth}
  \includegraphics[width=\linewidth]{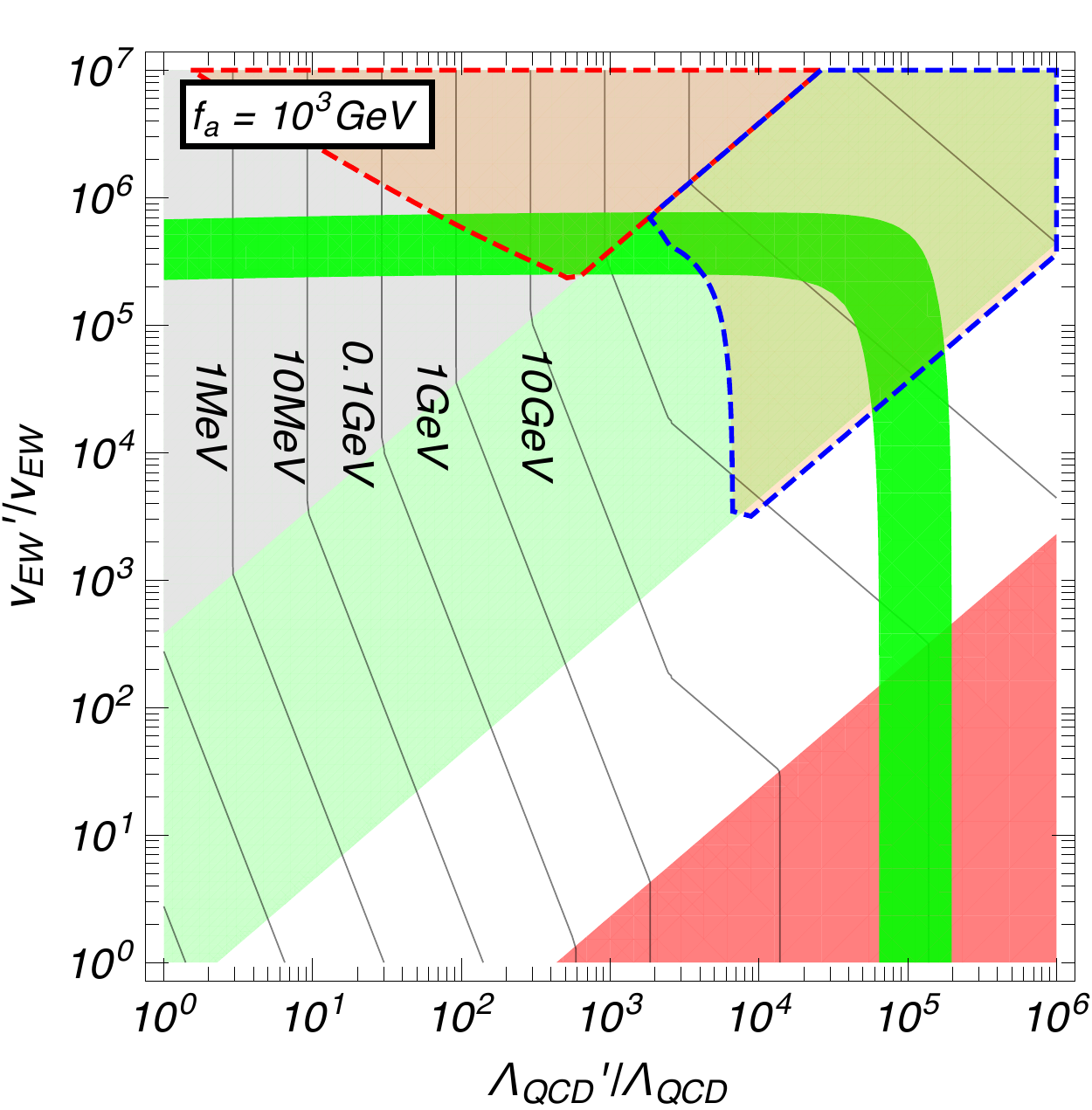}
 \end{minipage}
 \hspace{1cm}
 \begin{minipage}{.46\linewidth}
  \includegraphics[width=\linewidth]{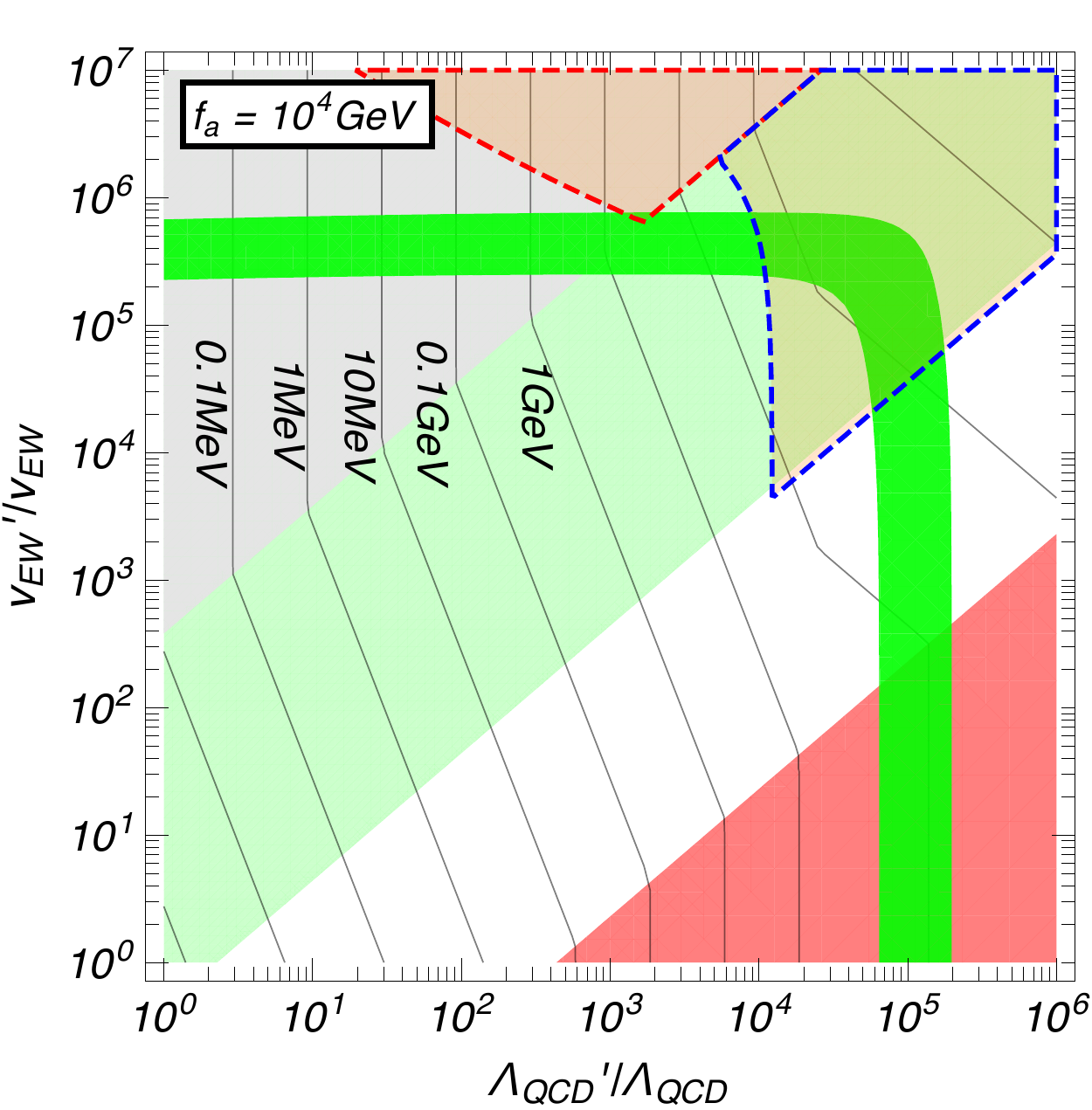}
 \end{minipage}

 \end{center}
\caption{\sl \small
The parameter region where the observed dark matter density is explained (green band)
for given  $f_a$ in the presence of a very light $\nu'$. There, the dominant components of the dark matter are $n'$, $e'$ and $e'$ and $\Delta'^{++}$ for the vertical, curved and horizontal regions, respectively.
The contour plot of the axion mass is also shown.
The gray, blue, red and green shaded regions are the same with the ones in Fig.\,\ref{fig:ma}.
The areas enclosed by the red and blue dashed lines are excluded by the 
the constraints on the dark matter annihilation from CMB observations.
}
\label{fig:relic2}
\end{figure}

In Fig.\,\ref{fig:relic2}, we show that parameter where the observed dark matter density is explained.
Here, we use the annihilation cross sections given in the previous section
and we again allow the predicted dark matter density within $\Omega h^2 = 0.03$--$0.3$.
In this case, the observed dark matter density can be explained by $N'$ with $m_{N}' \simeq 100$\,TeV.
In the heavy quark$'$ region,  dark matter consists of $\Delta'^{++}$ with a mass in the TeV range 
and $e'$ with a mass around $300$\,GeV.%
\footnote{Here, the resultant number density of $\Delta'^{++}$ after confinement is similar to that of $u'$.}

As a notable difference from the case with  $m_{\nu}' > m_{\pi^\pm}'$, there is a parameter region
where dark matter mainly consists of neutral particle $n'$ while $p'$ decays away.
Since $n'$ does not couple to a long range force, this parameter region is free from
the CMB constraints on the annihilation cross section at around the recombination time
as well as other constraints on the self-interactions of dark matter.

%%%%%%%%%%%%%%%%%%%%%%%%%%%%
%%%%%%%%%%%%%%%%%%%%%%%%%%%

Before closing this section, let us comment that  the CMB constraints on the annihilation cross section at around the recombination time
as well as other constraints on the self-interactions of dark matter
can be easily evaded if $\text{U}(1)_{\rm QED}'$ is spontaneously broken and $\gamma'$ obtains 
a finite mass.
Such spontaneous breaking is easily achieved when each sector has two Higgs doublets.
There, the $\text{U}(1)_{\rm QED}'$ can be broken with appropriate couplings between
the two Higgs doublets in the two sectors.
In this case, entire regions on the green band in Fig.\,\ref{fig:relic2} are 
viable to explain the observed dark matter density with no long-range interactions.%
\footnote{Here, we assume that $m_{\gamma}' < m_e'$ so that $e'$ can annihilate into $\gamma'$.
It is also noted that $\pi'^\pm$ decays into a pair of massive $\gamma'$'s.}

%%%%%%%%%%%%%%%%%%%%%%%%%%%%%%%%%%%%%%%%%%%%%%%%%%%%%%%%%%%%
\subsection{$\nu'$ Dark Matter}
As a final possibility, let us consider that Dirac $\nu'$ dark matter which is possible for $m_\nu' < m_{\pi^\pm}'$.
The annihilation cross section of $\nu'$ into a pair of $e'$, $\mu'$ and $\tau'$ via $Z'$ exchange 
is given by \cite{Lee:1977ua}
\begin{eqnarray}
\vev{\sigma v_{\rm rel}} \simeq \frac{3m_\nu'^2}{16\pi \cos^4\theta_W' v_{EW}'^4}
 \left(
\left(
\frac{1}{2} -\sin^2\theta_W'
\right)^2 + 
\left(
\frac{1}{2}
\right)^2
\right)\ ,
\end{eqnarray}
where $\theta_W'$ is the weak mixing angle in the mirrored sector.
Thus, the appropriate $\nu'$ dark matter density is obtained when the Dirac neutrino mass satisfies
\begin{eqnarray}
\label{eq:nu1}
m_{\nu}' = y_\nu' v_{EW}' \simeq 8\,{\rm GeV}\times \left( \frac{v_{EW}'}{v_{EW}}\right)^2\ .
\end{eqnarray}
Here, $y_{\nu}'$ denotes the neutrino Yukawa coupling in the mirrored sector, and we assume $\theta_W'\simeq  \theta_W$ in the final expression.

%%%%%%%%%%%%%%%%%%%%%%%%%%%%
\begin{figure}[t]
\begin{center}
 \begin{minipage}{.46\linewidth}
  \includegraphics[width=\linewidth]{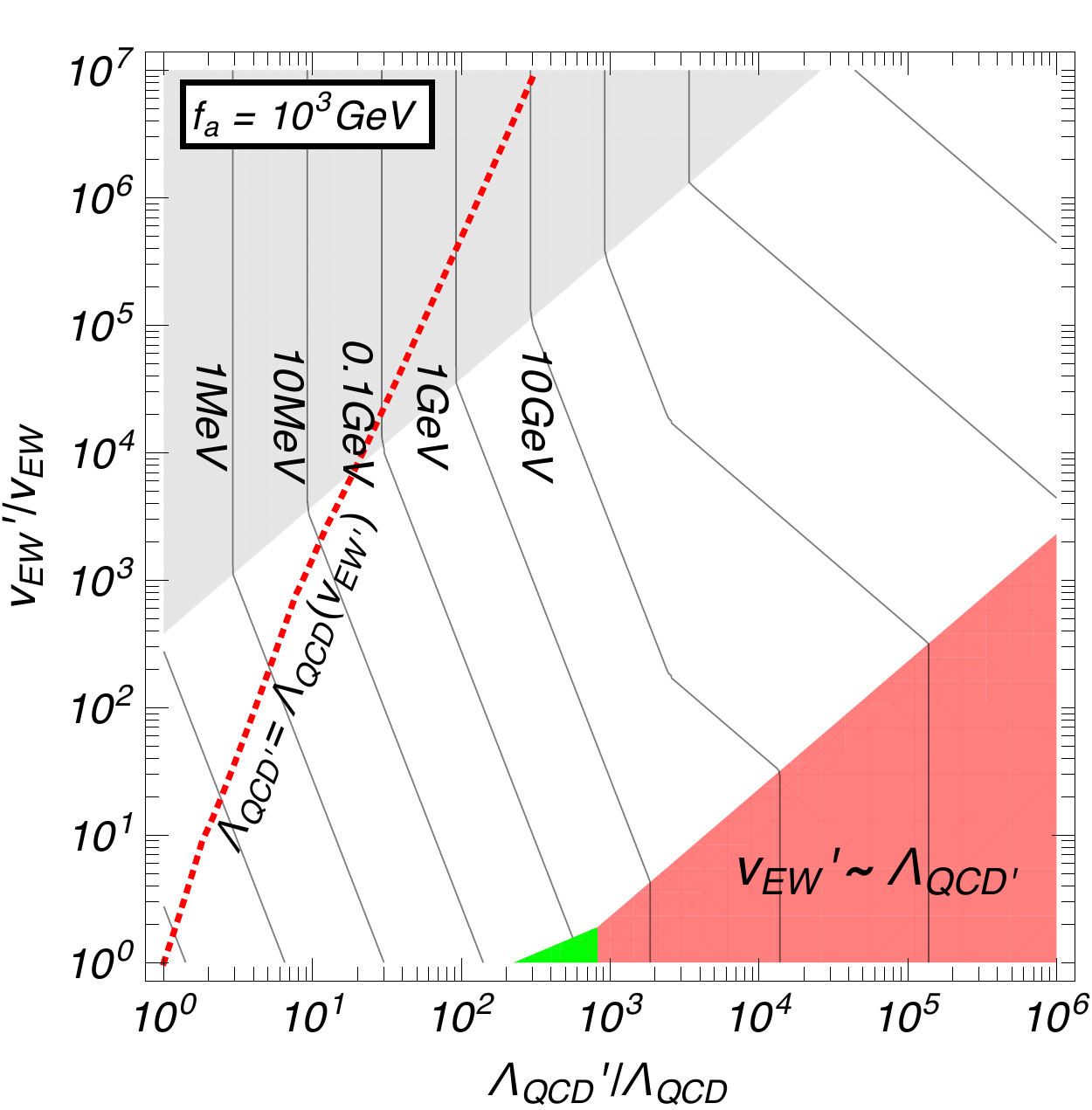}
 \end{minipage}
 \hspace{1cm}
 \begin{minipage}{.46\linewidth}
  \includegraphics[width=\linewidth]{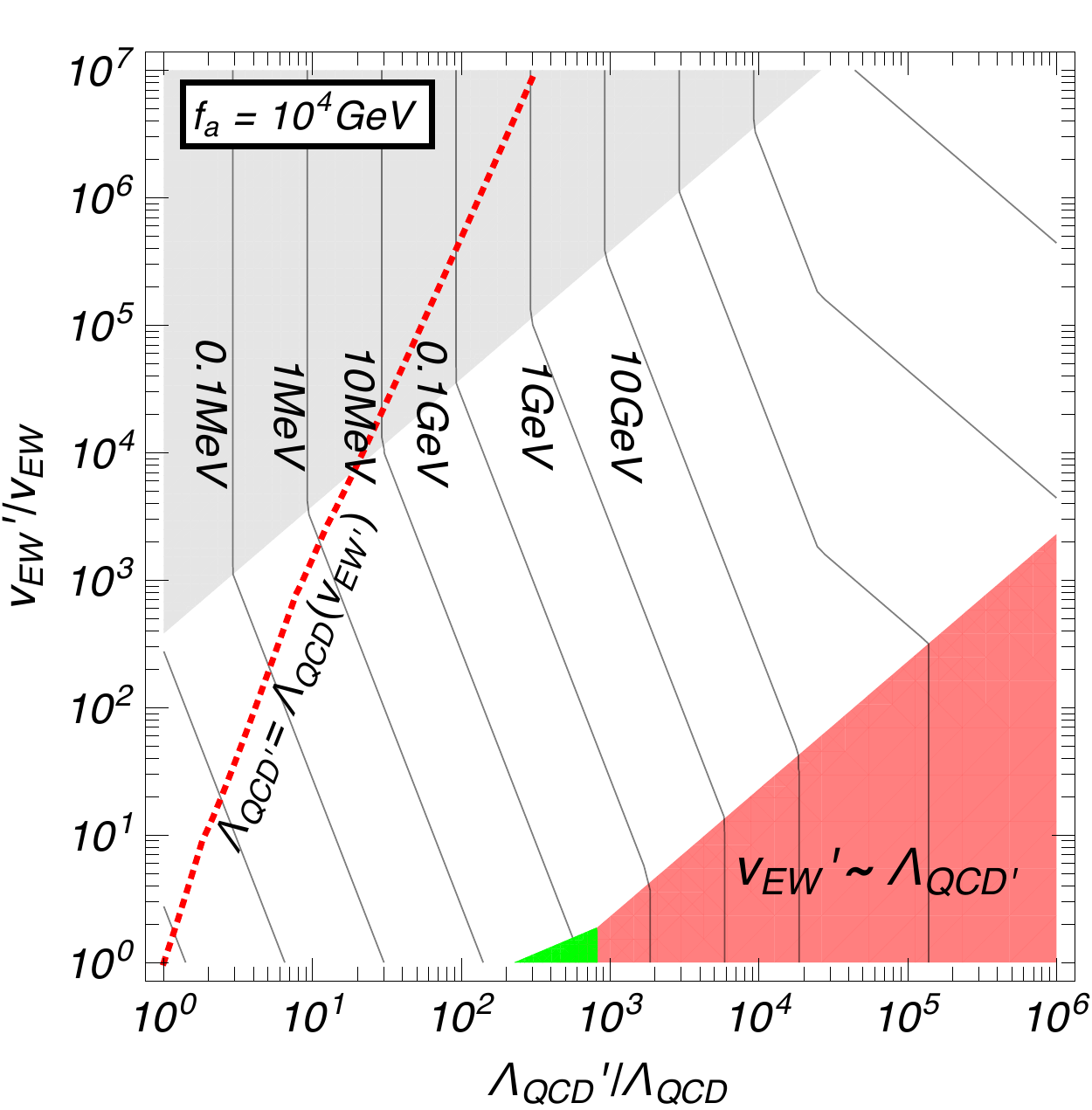}
 \end{minipage}
 \end{center}
\caption{\sl \small
The parameter region where the observed dark matter density is explained 
by $\nu'$ annihilating into leptons via $s$-channel $Z'$ boson. 
In the green shaded region, the condition, $m_\nu' < m_{\pi^\pm}'$ is satisfied, and hence,
$\nu'$ is stable.
}
\label{fig:nu1}
\end{figure}
%%%%%%%%%%%%%%%%%%%%%%%%%%%

In Fig.\,\ref{fig:nu1}, we show the parameter space which satisfies Eq.\,(\ref{eq:nu1}) and $m_{\nu}' < m_{\pi^\pm}'$.
The figure shows that only a small portion of the parameter space  is allowed.
As the figure shows, the corresponding axion mass is lighter than $100$\,MeV range for $f_a = 10^4$\,GeV
which are excluded by the beam dump experiments and cosmological arguments~\cite{Fukuda:2015ana}.
The axion mass for $f_a = 10^3$\,GeV is also close to the exclusion limits though not ruled out.

So far, we have assumed that the $B-L$ symmetry and  the $B'-L'$ symmetry are global symmetries
or at most discrete gauge symmetries which are not associated with gauge bosons.
If we consider that they are continuous gauge symmetries, on the other hand, $B'-L'$ gauge boson is in the mirrored 
sector is massless, and hence, $\nu'$s can annihilate into  $B'-L'$ gauge bosons
with an annihilation cross section,
\begin{eqnarray}
\vev{\sigma v_{\rm rel}} = \frac{\pi \alpha_{\rm B-L}'^2}{2m_{\nu}'^2}\ .
\end{eqnarray}
With this cross section, the dark matter density is explained for 
\begin{eqnarray}
m_\nu' \simeq 400\,{\rm GeV} \times \left(\frac{\alpha_{B-L}'}{10^{-2}}\right)\ ,
\end{eqnarray}
which can be consistent with $m_{\nu}' < m_{\pi^\pm}'$ in large parameter region.
Furthermore, by allowing  slight spontaneous breaking, the constraints on the 
Sommerfeld enhanced annihilation as well as  other constraints on the self-interactions of dark matter
can be evaded.

\section{Conclusions and Discussions}
\label{sec:conclusions}
In this paper, we discussed dark matter candidates in the visible heavy QCD axion model.
As we have shown, $\pi'^\pm$ and $e'^\pm$ 
can be a viable candidate for dark matter when it is lighter than all of $\nu'$
for $f_a = 10^{3}$--$10^4$\,GeV.
As an interesting feature, they serve as self-interacting dark matter with a long range force.
We also showed  $n'$ can be also a viable dark matter candidate when 
its mass is around $100$\,TeV with one of $\nu'$ being very light or massless.
It is also shown that $\nu'$ can also be a viable candidate for dark matter.
In particular, we find that $\nu'$ can be viable candidate in a large parameter region
when the $B'-L'$ gauge interaction is invoked.

For a moderate value of the decay constant, $f_a \lesssim 10^4$\,GeV, the model can be tested at future collider experiments via 
the direct production of $s$, $a$, and the extra quarks required for the PQ-mechanism.
Besides, the darkly-charged dark matter candidates annihilating into $\pi'^0$ leave imprints on the spectrum the CMB anisotropy
through the $a$--$\pi'^0$ mixing (see Fig.\,\ref{fig:relic1}).
The future CMB observations such as {PIXIE}~\cite{Kogut:2011xw}
{LiteBird}~\cite{Matsumura:2013aja}, and {CORE$+$}~\cite{Martins:2015dtz} will be able to improve the limit on the annihilation 
cross section at around the recombination time.
The darkly-charged dark matter candidates can also be strengthen if future observations of dark halo structure reveal 
that dark matter should have a long-range force.

Another dark matter candidate, $n'$ in the hundreds TeV range, also annihilates into the axion through  the $a$--$\pi'^0$ mixing.
By assuming the total annihilation cross section in Eq.\,(\ref{eq:WIMP}),
the annihilation cross section into the axion is of ${\cal O}(10^{-28})$\,cm$^3/$s. 
Such a cross section  is much lower than the current 
constraints from the antiproton to proton ratio in the cosmic ray~\cite{Giesen:2015ufa,Ibe:2015tma} measured
by AMS-02~\cite{AMS02:2015}.%
\footnote{Here, we roughly translate the constraints in \cite{Giesen:2015ufa,Ibe:2015tma} 
for the dark matter model annihilating into $b\bar{b}$ and $W^+W^-$ for $M_a \gtrsim {\cal O}(1)$\,GeV.
For a lighter axion, it does not lead to anti-proton signals, and hence, the constraints are much weaker.}
It is also lower than the constraints from the continuous gamma ray spectrum from the dwarf spheroidal galaxies
measured by Fermi-LAT~\cite{Ackermann:2015zua}.

Finally, let us consider  the ``nucleon$'$ decay"
as an intriguing probe of the $n'$ dark matter candidate in the hundreds TeV range. 
Since the $B$ and $B'$ symmetries are global symmetries, they are expected to be broken at least
by Planck suppressed operators as generically expected  in quantum gravity.
Thus, through the Planck suppressed dimension six operators for example, 
the decay rate of $n'$ into $\nu'$ and $\pi'^0$ is roughly given by
\begin{eqnarray}
\Gamma(n' \to \nu' + \pi'^0) \sim \frac{1}{32\pi} \frac{m_{N'}^5}{M_{\rm PL}^4} \ ,
\end{eqnarray}
where $M_{\rm PL} \simeq 2.4 \times 10^{18}$\,GeV is the reduced Planck scale.%
\footnote{For a rough estimation, we neglect uncertainties in hadronic matrix elements.}
A fraction of  $n'$ decays also into axion through the $a$--$\pi'^0$ mixing of ${\cal O}(f_\pi'/f_a)$,
which subsequently decays into the QCD jets.
Altoghether, the lifetime of $n'$ divided by the branching ratio into the axion is roughly given by,
\begin{eqnarray}
\label{eq:nprime}
\tau(n' \to \nu' + a) \sim 10^{28}\,{\rm s}\times \left(\frac{100\,{\rm TeV}}{m_{N'}}\right)^5
\left(\frac{f_a}{100\,\rm TeV}\right)^2
\left(\frac{10\,\rm TeV}{f_\pi'}\right)^2\ .
\end{eqnarray}

The decay of dark matter into QCD jets is constrained from the observations of the extragalactic 
gamma-ray background (EGRB)~\cite{Ibarra:2007wg,Ishiwata:2009dk,Carquin:2015uma,Ando:2015qda,*Ando:2016ang}.
The constraint on the lifetime of $n'$ decaying into QCD jet can be read from \cite{Ando:2015qda,*Ando:2016ang}
\begin{eqnarray}
\label{eq:lifetime}
\tau(n' \to \nu' + a) \gtrsim 10^{28}\,{\rm s}\times \left(\frac{\Omega_{n'}}{\Omega_{DM}} \right)
\ .
\end{eqnarray}
Notably, the constraint from the EGRB observations is close to the lifetime (divided by the branching ratio into the axion) in Eq.\,(\ref{eq:nprime})
for $\Omega_{n'} = \Omega_{DM}$.
Therefore, the EGRB observations are indirectly probing the global symmetry breaking expected in quantum gravity 
through the $n'$ decay in the mirrored sector.

Furthermore,  $n'$ dark matter can also be tested by the proton decay searches in the Standard Model sector
if the Grand Unified Theory (GUT) exists at a scale $M_{\rm GUT}$ lower than the Planck scale.
Under the assumption of the GUT, two sectors are expected to have the same GUT scale, $M_{\rm GUT}$, 
due to the $\mathbb Z_2$ exchanging symmetry.
Therefore, the $n'$ lifetime divided by the branching ratio into the axion is roughly interrelated to the proton lifetime $\tau_p$ 
in the Standard Model sector,
\begin{eqnarray}
\tau_p(p\to e+\pi^0) \simeq 10^{35}\, {\rm yr}\times \left(\frac{\alpha_{\rm GUT}^{-1}}{25}\right)^2
\left(\frac{M_{\rm GUT}}{10^{16}\,\rm GeV}\right)^4\ ,
\end{eqnarray}
as
\begin{eqnarray}
\label{eq:lifetime2}
\tau(n' \to \nu' + a) \sim 3 \times10^{19}\,{\rm s}\times 
\left(\frac{\tau_p}{10^{35}\,\rm yr}\right)
\left(\frac{100\,{\rm TeV}}{m_{N'}}\right)^5
\left(\frac{f_a}{100\,\rm TeV}\right)^2
\left(\frac{10\,\rm TeV}{f_\pi'}\right)^2\ .
\end{eqnarray}
Here $\alpha_{GUT}$ denotes the fine-structure constant of the Grand Unified Theory.
Thus, if the Hyper-Kamiokande experiment observes the proton decay with a lifetime of ${\cal O}(10^{35})$\,yr~\cite{Hyper-Kamiokande:2016dsw}, 
the $n'$ dark matter candidate is immediately excluded in combination with the EGRB observation in Eq.\,(\ref{eq:lifetime}).

%%%%%%%%%%%%%%%%%%%%%%%%%%%%%%%%%%%%%%
%%%%%%%%%%%%%%%%%%%%%%%%%%%%%%%%%%%%%%
\section*{Acknowledgements}
The authors thank Cheng-Wei Chiang for useful discussions  at the early stage of the project.
This work is supported in part by Grants-in-Aid for Scientific Research from the Ministry of Education, Culture, Sports, Science, and Technology (MEXT) KAKENHI, 
Japan, No.\,25105011 and No.\,15H05889 (M. I.) as well as No.\,26104009 (T. T. Y.); Grant-in-Aid No.\,26287039 (M. I. and T. T. Y.) and  No.\,16H02176 (T. T. Y.) 
 from the Japan Society for the Promotion of Science (JSPS) KAKENHI; and by the World Premier International Research Center Initiative (WPI), MEXT, Japan (M. I., and T. T. Y.).
 The work of H.F. is supported in part by a Research Fellowship for
 Young Scientists from the Japan Society for the Promotion of Science (JSPS).
%%%%%%%%%%%%%%%%%%%%%%%%%%%%%%%%%%%%%%

\appendix
\section{Explicit Breaking of the PQ-Symmetry}
\label{sec:breaking}
Throughout this paper, $\text{U}(1)$ PQ-symmetry is assumed to be an almost exact symmetry of the model
broken only by the axial anomaly.
It is believed, however, that global symmetries are to be broken by Planck suppressed operators as generically expected  in quantum gravity.
For example, Planck suppressed  self-interacting operators of $\phi$
\begin{eqnarray}\label{eq:QG}
{\cal L}_{\cancel{PQ}} = \frac{\k}{(n+4)!M_{\rm PL}^{n}} \left(\phi^{n+4} + \phi^{*n+4}\right)\ ,\quad (n> 0) \ ,
\end{eqnarray}
with $\k = O(1)$ break the PQ-symmetry explicitly.
They lead to a non-vanishing effective $\theta$-angle at the minimum of the axion potential, 
\begin{eqnarray}
{\mit\Delta}\theta_{\rm eff} \sim \frac{\kappa}{2^{(n+2)/2} (n+3)!}\frac{f_a^{n+2}}{M_{\rm PL}^n M_a^2}\ .
\end{eqnarray}
Thus, for dimension five operators ($n=1$), for example, the effective $\theta$-angle is given by
\begin{eqnarray}
{\mit \Delta} \theta_{\rm eff} \sim 10^{-10} \times\k\left(\frac{f_a}{10^{4}\,\rm GeV}\right)^{3}\left(\frac{10\,\rm GeV}{M_a}\right)^2\ ,
\end{eqnarray}
which is consistent with the current upper bound on the effective $\theta$-angle of ${\cal O}(10^{-11})$ for $f_a \lesssim O(10^3$--$10^4)$ and $M_a =O(0.1$--$10)$\,GeV.%
\footnote{This feature is also advantageous to make a model where the PQ symmetry appears as an accidental 
symmetry resulting from other exact gauge symmetries~(see \cite{Harigaya:2013vja,Redi:2016esr} and references therein).%
}

In addition to the self-interacting operators in Eq.\,(\ref{eq:QG}), the other types of operators such as
\begin{eqnarray}
\label{eq:mixPQ}
{\cal L}_{\cancel{PQ}} = \frac{|\Phi_{B-L}|^{2n}}{M_{\rm PL}^{2n-3}} \phi + h.c. \ ,
\end{eqnarray}
also leads to explicit breaking the PQ-symmetry. 
Here, $\vev{\Phi_{B-L}}$ is the order parameter of the $B-L$ symmetry.
If we assume $\vev{\Phi_{B-L}} \simeq 10^{10}$\,GeV, for example, 
the PQ-symmetry is badly broken for $n=2$.
To avoid this problem, it is required  to assume that  $\phi$ has
sizable couplings to the $\psi^{(\prime)}$, while it has highly suppressed couplings to the fields in the Standard Model sector and in the mirrored sector.

As another way to evade this problem, we may consider a model with an exact (and hence gauged) discrete symmetry under which $\phi$ rotates non-trivially.
For example, a model with a ${\mathbb Z}_5$ discrete symmetry can be constructed 
by introducing five pairs of $(\psi_L^{(\prime)},\bar\psi_R^{(\prime)})$. 
Under the ${\mathbb Z}_5$ symmetry, $\phi$ has a charge $-1$ while $\psi_L$ and $\psi'_L$ have the $\text{U}(1)$ charge $0$, 
and $\bar \psi_R$ and $\bar\psi'_R$ have the $\text{U}(1)$ charge $-1$, with which the discrete symmetry is free from anomalies.
In this model, the PQ-symmetry is realized an accidental symmetry while the PQ-breaking operators in Eq.\,(\ref{eq:mixPQ}) is forbidden.

One problem of the model with an exact discrete symmetry is that the model causes the domain wall problem
when it is spontaneously broken by $\vev\phi$~\cite{Zeldovich:1974uw,Kibble:1976sj}.
This problem can be avoided by assuming that ${\mathbb Z}_5$ is embedded in a gauge $\text{U}(1)$ symmetry
so that $\text{U}(1)$ symmetry is broken at a scale not very higher than $f_a$.\,\footnote{In such embedded models, the Peccei-Quinn breaking operators like Eq.\,\ref{eq:mixPQ} never appear due to the gauge symmetry. Thus, not $\mathbb{Z}_5$ but $\mathbb{Z}_4$ remnant symmetry may be sufficient.}
For example, we may consider a $\text{U}(1)$ gauge symmetry under which 
$\phi$ has a charge $1$ while $\psi_L$ and $\psi'_L$ have a charge $0$, and $\bar \psi_R$ and $\bar\psi'_R$
have a charge $-1$.
Besides, we also introduce a scalar field $X$ with a $\text{U}(1)$ charge $-5$ and pairs of colored left-handed Weyl fermions 
$(\xi_L,\bar\xi_R)$ and $(\xi_L',\bar\xi_R')$ with $\xi^{(')}_L$ and 
$\bar\xi_R^{(')}$ having the $\text{U}(1)$ charges $3$ and $2$, respectively.%
\footnote{With this charge assignment, $\text{U}(1)$ gauge symmetry is free from anomalies except for $\text{U}(1)^3$ anomaly.
The $\text{U}(1)^3$ anomaly can be cancelled by introducing appropriate number of $\text{U}(1)$ charged fermions 
which are neutral under the Standard Model$^{(\prime)}$  gauge symmetries.}
Under the $\text{U}(1)$ gauge symmetry, $\psi$'s and $\xi$'s can couple via
\begin{eqnarray}
{\cal L} = \phi \psi_L^{(\prime)}\bar\psi_R^{(\prime)}
+ X \xi_L^{(\prime)}\bar\xi_R^{(\prime)} + h.c. \  .
\end{eqnarray}
Then, once $X$ obtains a VEV, the desired ${\mathbb Z}_5$ symmetry remains with which 
the PQ-symmetry is realized as an approximate approximate symmetry.
In this model, the domain wall is not stable and the domain wall problem can be evaded~\cite{InPrep}.

\bibliography{./papers}
%\bibliography{MirrorDM.bbl}

\end{document}